\RequirePackage{fix-cm}

\documentclass[smallextended]{svjour3}
\smartqed

\usepackage{amssymb,graphicx,afterpage,mathptmx}

\journalname{General Relativity and Gravitation}

\begin{document}

  \title{Neutrino trapping in braneworld extremely compact stars}

  \author{Zden\v{e}k Stuchl\'{\i}k \and
          Jan Hlad\'{\i}k          \and\\
          Martin Urbanec}

  \institute{Institute of Physics, Faculty of Philosophy and Science, Silesian University in Opava, Bezru\v{c}ovo n\'{a}m. 13, CZ-746\,01 Opava, Czech Republic\\
  \email{zdenek.stuchlik@fpf.slu.cz \and jan.hladik@fpf.slu.cz \and martin.urbanec@fpf.slu.cz}}

  \date{Received: date / Accepted: date}

  \maketitle

  \begin{abstract}
    Extremely Compact Stars (ECS) contain trapped null geodesics. When such objects enter the evolution period admitting geodetical motion of neutrinos, certain part of neutrinos produced in their interior will be trapped influencing their neutrino luminosity and thermal evolution. We study neutrino trapping in the braneworld ECS, assuming uniform distribution of neutrino emissivity and massless neutrinos. We give the efficiency of the neutrino trapping effects in the framework of the simple model of the internal spacetime with uniform distribution of energy density, and external spacetime described by the Reissner-Nordstr\"{o}m geometry characterized by the braneworld ``tidal'' parameter $b$. For $b < 0$ the external spacetime is of the black-hole type, while for $b > 0$ the external spacetime can be of both black-hole and naked-singularity type. Then the ECS surface radius $R$ can be located also above the unstable (outer) photon circular orbit. Such basically new types of the spacetimes strongly alter the trapping phenomena as compared to the standard case of $b = 0$. It is shown that the neutrino trapping effects are slightly lowered by the presence of physically more plausible case of $b < 0$, as compared to the standard internal Schwarzschild spacetime, while they can be magnified by positive tidal charges if $b < 1$ and lowered for $b > 1$. However, potential astrophysical relevance of the trapping phenomena is strongly enhanced for negative tidal charges enabling a significant enlargement of the ECS surface radius to values coherent with recent observations.

    \keywords{Neutrino trapping \and Braneworlds \and Extremely compact stars}
    \PACS{95.30.Sf \and 04.50.Kd}
  \end{abstract}

  \section{Introduction}
    One of the promising approaches to the higher-dimensional gravity theories is represented by the braneworld models, where the observable universe is a 3-brane (domain wall) to which the matter fields are confined, while the gravity field enters the extra spatial dimensions of size that could strongly exceed the Planck length scale \cite{Ark-Dim-Dva:1998:}. Such models provide an elegant solution to the hierarchy problem of electroweak and quantum gravity scales, as these become of the same order ($\sim \mathrm{TeV}$) due to large extra dimensions and could be well tested on the planned supercollider experiments \cite{Dim-Lan:2001:}. Gravity can be localized near the brane at low energies even with an infinite size extra dimension with warped spacetime satisfying the 5D Einstein equations containing negative cosmological constant \cite{Ran-Sun:1999:} and an arbitrary energy-momentum tensor allowed on the brane \cite{Shi-Mae-Sas:1999:}. The Randall-Sundrum model gives 4D Einstein gravity in low energies and the Newtonian limit appears on the 3-brane with high accuracy. Significant deviations from the Einstein gravity occur at very high energies, in the early universe and in vicinity of compact objects as black holes and neutron stars \cite{Maar:2004:}.

    Recently, no exact solution of the full 5D Einstein equations is know, but there is a variety of astrophysically plausible special solutions of the 4D effective Einstein equations constrained to the brane. Such solutions describe black holes with spherical symmetry \cite{Dad-Maar-Pap-Rez:2000:PHYSR4:} or with axial symmetry \cite{Ali-Gum:2005:} and compact objects that could represent neutron (quark) stars \cite{Ger-Maar:2001:}. The black hole spacetimes are determined by geometry of the spherical Reissner-Nordstr\"{o}m (R-N) and axial Kerr-Newman (K-N) type where the electric charge squared is substituted by the braneworld ``tidal charge'' parameter representing the tidal (Weyl tensor) effects of the bulk space onto the 4D black hole structure. Astrophysically relevant properties of the braneworld black holes were studied in a series of papers devoted to both motion of matter in their vicinity \cite{Stu-Kot:2009:,Ali-Tal:2009:,Abdu-Ahme:2010:PHYSR4,Mam-Hak-Toj:2010:MPLA:,Mor-Ahme-Abdu-Mam:2010:ASS:} and optical phenomena \cite{Sche-Stu:2009:a,Sche-Stu:2009:b,Bin-Nun:2010:PHYSR4:a,Bin-Nun:2010:PHYSR4:b}.

    In the simple model of spherically symmetric stars with uniform energy density profile a variety of special solutions having asymptotically Schwarz\-schil\-di\-an character and satisfying the braneworld boundary conditions were found \cite{Ger-Maar:2001:}. The most popular is the one with external spacetime described by the Reissner-Nordstr\"{o}m geometry with the braneworld tidal charge parameter reflecting the tidal effects of the bulk and related to the energy density and brane tension. Its properties were extensively studied both in the weak field limit \cite{Boh-Har-Lob:2008:CLAQG:,Boh-Ris-Har-Lob:2010:CLASQG} and strong field limit when some restrictions on the brane tension were implied from the data of kHz~QPOs observed in low mass binary systems with a neutron star \cite{Kot-Stu-Tor:2008:CLASQG:}. Here we focus our attention to the interior of the compact objects, namely to the phenomena related to trapping of neutrinos in the so called Extremely Compact Stars (ECS) admitting (by definition) existence of trapped null geodesics \cite{Stu-Tor-Hle-Urb:2009:}. Usually, the braneworld tension is assumed positive while the related tidal charge has to be negative \cite{Ger-Maar:2001:}, but the negative tension and related positive tidal charge are not excluded \cite{Dad-Maar-Pap-Rez:2000:PHYSR4:}, so we consider here both positive and negative tidal charge of the compact object. Note that exterior of neutron stars with positive tidal  charges can be described by both black-hole and naked-singularity types of the external R-N spacetime (for details see \cite{Kot-Stu-Tor:2008:CLASQG:}).

    In the standard ($b = 0$) internal Schwarzschild spacetimes of uniform energy density \cite{Schw:1916:SITBA:,Stu:2000:ACTPS2:} with radius $R < 3GM/c^2$, bound null geodesics exist being concentrated around the stable circular null geodesic~\cite{Abr-Mil-Stu:1993:PHYSR4:,Stu-etal:2001:PHYSR4:}. From the behaviour of the effective potential of null geodesics in the exterior, vacuum Schwarzschild spacetimes, determining the unstable null circular geodesics at the radius $r_{\mathrm{ph}} = 3GM/c^2$ (see, e.g., \cite{Mis-Tho-Whe:1973:Gra:}), we can conclude that any spherically symmetric, static non-singular interior spacetime with radius $R < r_{\mathrm{ph}}$ admits existence of bound null geodesics. The trapped null geodesics are then governed by both internal and external barriers of the null geodesics effective potential. \footnote{In principle, the bound null geodesics could exist also in objects having $R > 3GM/c^2$, e.g., in some composite polytropic spheres \cite{Nil-Cla:2000:GRRelStarsPolyEOS:}.} The realistic equations of state admitting the existence of the extremely compact objects were found and investigated for neutron stars, quark stars and Q-stars \cite{Bah-Lyn-Sel:1990:ApJ:,Mil-Sha-Nol:1998:MNRAS:,Nil-Cla:2000:GRRelStarsPolyEOS:,Stu-Tor-Hle-Urb:2009:,Hle-Stu-Mra:2004:RAGtime4and5:CrossRef,Oest:2001:RAGtime2and3:}. For spacetimes with a non-zero tidal charge, the situation is more complex as we shall demonstrate in the following.

    The existence of bound null geodesics in ECS has interesting astrophysical consequences. For example, using the notion of the optical reference geometry \cite{Abr-Pra:1990:,Stu-Hle-Jur:2000:,Stu:1990:} it was shown that trapped modes of gravitational waves could influence some instabilities in these objects \cite{Abr-etal:1997:CLAQG:}. Trapping of neutrinos can be of high importance in the interior of the ECS: first, it will suppress the neutrino flow as measured by distant observers and, second, it can influence cooling of the ECS in a layer extending from some radius depending on details of their structure up to the surface radius. The cooling process could even be realized in a ``two-temperature'' regime, when the temperature profile in the interior of the star with no trapped neutrinos differs from the profile established in the external layer with trapped neutrinos \cite{Stu-Tor-Hle-Urb:2009:} modifying substantially the standard picture of the neutron star structure as given in \cite{Gle:1992:PHYSR4:,Gle:2000:CompactStars:,Hae-Zdu:1986:NATURE:,Web-Gle:1992:ASTRJ2:,Web:1999:Pul:}. In a different context, trapping of neutrinos inside rotating neutron stars has been discussed in \cite{Mallick-etal:2009:arXiv:0905.3605:}.

    The approximation of free, geodetical motion of neutrinos in the internal spacetime could be used when the mean free path of neutrinos $\lambda > R$. Neutrinos have inelastic scatter on electrons (muons) and elastic scatter on neutrons. The scatter cross section on electrons (neutrons) $\sigma_\mathrm{e}$ ($\sigma_\mathrm{n}$) determines the mean free path by formula $\lambda = \left(\sigma_{i} n_{i}\right)^{-1}$ where $n_i$~($i = \mathrm{e, n}$) denotes the number density of electrons (neutrons). It was shown \cite{Sha-Teu:1983:BHWDNS:} that
    \begin{equation}
      \lambda_\mathrm{e}\sim 9\times 10^7\left(\frac{\rho_\mathrm{nucl}}{\rho}    \right)^{4/3}\left(\frac{100~\mathrm{keV}}{E_\nu}\right)^{3}~\mathrm{km},
    \end{equation}
    while
    \begin{equation}
      \lambda_\mathrm{n}\sim 300\frac{\rho_\mathrm{nucl}}{\rho}    \left(\frac{100\mathrm{keV}}{E_\nu}\right)^{2}~\mathrm{km}.
    \end{equation}
    There is $\lambda_\mathrm{e} \gtrsim  10$~km for $E_\nu \lesssim 20$~MeV and $\lambda_\mathrm{n} \gtrsim 10$~km for $E_\nu \lesssim 500$~keV. Therefore, in a few hours old neutron star, see \cite{Lat-Pra:2007:PhysRep:,Sha-Teu:1983:BHWDNS:}, at temperatures $T \lesssim 10^9$~K ($E_\nu \sim 100$~keV), the neutrino motion could be considered geodetical through whole the internal spacetime.

    Bound neutrinos with mean free path${}\gg R$ will slow down the cooling. Of course, they will be re-scattered due to finiteness of the mean free path. An eventual scattering of trapped neutrinos will cause change of their impact parameter, therefore, some of them will escape the ECS, suppressing thus the slow down of the cooling process in the region of neutrino trapping. Clearly, the scattering effect of the trapped neutrinos is a complex process deserving sophisticated numerical code based on the Monte Carlo method (we expect modelling of this effect in future).

    The cooling process deserves sophisticated analytical estimates and detailed numerical simulations. As a first step in considering the role of trapped neutrinos in ECS, efficiency of the neutrino trapping effect was studied in the simple case of the internal Schwarzschild spacetime with uniform distribution of energy density (but a nontrivial pressure profile) and isotropic and uniform distribution of local neutrino luminosity, when all the calculations can be realized in terms of elementary functions only (see \cite{Stu-Tor-Hle-Urb:2009:}). \footnote{The internal Schwarzschild spacetime can well represent the spacetime properties of realistic ECS~\cite{Gle:1992:PHYSR4:,Gle:2000:CompactStars:,Hae-Zdu:1986:NATURE:} which are crucial in estimating the role of neutrino trapping. Such a simple spacetime could serve quite well as a test bed for realistic models of ECS.}
    The influence on the neutrino luminosity of the star is given by a luminosity trapping coefficient relating the total number of trapped neutrinos and the total number of radiated neutrinos (per unit time of distant observers). The influence on the cooling process is given by two ``cooling'' trapping coefficients: a ``local'' one given by ratio of trapped and radiated neutrinos at any radius where the trapping occurs, and the ``global'' one giving ratio of trapped and radiated neutrinos (per unit time of distant observers) integrated over whole the region where the trapping occurs.

    In the present paper we study the role of the braneworld ``tidal charge'' parameter on the existence of ECS and estimate its role in the efficiency of the neutrino trapping process using the same trapping coefficients as those introduced in~\cite{Stu-Tor-Hle-Urb:2009:}. We consider the trapping effect in the simple case of objects with uniform distribution of energy density. The external field is given by the Reissner-Nordstr\"{o}m geometry \cite{Ger-Maar:2001:}. This special solution is most extensively studied in the literature, since its external field is identical to those of the black hole solution with spherical symmetry \cite{Dad-Maar-Pap-Rez:2000:PHYSR4:} --- therefore, the results of our study could be efficiently compared to the results obtained in studies of other phenomena \cite{Boh-Har-Lob:2008:CLAQG:,Kot-Stu-Tor:2008:CLASQG:}. We consider both positive and negative values of the braneworld tidal parameter --- usually, only the negatively valued tidal charges (corresponding to positive tension of the brane) are discussed \cite{Ger-Maar:2001:}, but the inverse situation with positive tidal charge is not excluded and will be treated here, since it could demonstrate some new interesting effects.

    Our paper is organized as follows. In Section~\ref{SECbrwns}, we summarize properties and matching conditions of the internal uniform energy density spacetime and the Reiss\-ner-Nordstr\"{o}m external spacetime. In Section~\ref{SECogeoi}, null geodesics of both the internal and external spacetime are described in terms of properly given effective potential and the ECS are classified according to the properties of the trapping region of the null geodesic motion. In Section~\ref{SECtrappn}, the trapping of neutrinos is discused. In Section~\ref{SECeffnt}, the trapping efficiency coefficients are defined for both the total neutrino luminosity and neutrino cooling process, and determined for all kinds of the braneworld ECS. In Section~\ref{SECconcl}, concluding remarks are presented. Throughout the paper, we shall use the high-energy units with $\hbar = c = k_\mathrm{B} = 1$, if not stated otherwise. For simplicity, we assume zero rest energy of neutrinos, isotropic emission at each radius, and the period of evolution of the compact stars, when the temperature is low enough to describe the motion of neutrinos by the null geodesics of the spacetime.

  \section{Braneworld neutron stars}\label{SECbrwns}
    We describe the braneworld neutron stars by the simple model of uniform energy density interior and the ``tidally'' charged exterior. We give the internal and external geometry and the matching conditions. Accordingly, we determine
    conditions on the existence of ECS depending on the tidal parameter characterizing the external spacetime.

  \subsection{Matching conditions of internal and external spacetimes}
    In the standard Schwarzschild coordinates and the high-energy units, the line element of the
    spherically symmetric spacetimes reads

    \begin{equation}
      \mathrm{d}s^2 = -A^{2}(r)\mathrm{d}t^2 + B^{2}(r)\mathrm{d}r^2 + r^{2}\mathrm{d}\Omega.\label{EQmetric}
    \end{equation}

    The internal solution is characterized by the uniform energy density distribution
    --- $\varrho = \mathrm{const}$, and by the tension of the brane --- $\lambda$. We assume $\varrho > 0$, but we allow both possibilities $\lambda > 0$, $\lambda < 0$ for completeness. The internal geometry is matched to the external geometry at the surface of the star $r = R$. The line element of the internal solution is given by the metric coefficients $A^{-}(r)$, $B^{-}(r)$ that are determined by \cite{Ger-Maar:2001:}

    \begin{equation}
      A^{-}(r) = \frac{\Delta(R)}{\left(1+p(r)/\varrho\right)}\label{EQamin}
    \end{equation}
    and
    \begin{equation}
      \left(B^{-}(r)\right)^2 = \frac{1}{\Delta^{2}(r)} = \left[1-\frac{2G M}{r}\left(\frac{r}{R}\right)^3 \left(1+\frac{\varrho}{2\lambda}\right)\right]^{-1},
    \end{equation}
    where $M = \frac{4}{3}\pi\varrho R^3$.

    The pressure radial profile is given by
    \begin{equation}
      \frac{p(r)}{\varrho} = \frac{\left[\Delta(r) - \Delta(R)\right]\left(1 + \varrho/\lambda\right)}{3\Delta(R) - \Delta(r) + \left[3\Delta(R) - 2\Delta(r)\right]\left(\varrho/\lambda\right)}.
    \end{equation}
    For $r = R$, there is $p(r)/\varrho = 0$. The maximum of the pressure profile is at $r = 0$.

    The reality condition on the metric coefficient $B^{-}(r)$ (taken at $r = R$) implies a relation between $\lambda$, $\varrho$ and $R$ that can be expressed in the form
    \begin{equation}\label{EQRm2GM}
      \frac{G M}{R-2G M} \geq \frac{\varrho}{\lambda}.
    \end{equation}

    Considering the restriction $R > 2 G M$ ($R < 2 G M$), we can see that the reality condition~(\ref{EQRm2GM}) is satisfied for all $\lambda < 0$ (forbidden for all $\lambda > 0$), while for positive tension $\lambda > 0$ (negative tension $\lambda < 0$), we obtain a limit on the positive (negative) tension given by~\cite{Ger-Maar:2001:}
    \begin{equation}
      \lambda \geq \left(\frac{R - 2 G M}{G M}\right)\varrho .
    \end{equation}
    (Notice that considering possibility of $R < 2 G M$, we obtain a limit on the negative tension.)

    The line element of the external geometry is given by the metric coefficients $A^{+}(r)$, $B^{+}(r)$ that are determined by
    \begin{equation}
      \left( A^{+}(r)\right)^{2} = \left( B^{+}(r)\right)^{-2} = 1 - \frac{2G\mathcal{M}}{r} + \frac{q}{r^2},
    \end{equation}
    where, due to the matching conditions on the neutron star surface, i.e., $A^{-}(R) = A^{+}(R)$, $B^{-}(R) = B^{+}(R)$, the external mass parameter $\mathcal{M}$ and external tidal charge parameter $q$ are  related to the internal geometry parameters $\varrho$, $\lambda$ (and $M$) by the relations

    \begin{equation}
      q = -3G M R\frac{\varrho}{\lambda}
    \end{equation}
    \begin{equation}
      \mathcal{M} = M \left(1-\frac{\varrho}{\lambda}\right).
    \end{equation}

    For $\lambda > 0$, the tidal charge $q < 0$ and $\mathcal{M} < M$, while for $\lambda < 0$, there is $q > 0$ and $\mathcal{M} > M$. Notice that for $\lambda > 0$, the condition $\varrho < \lambda$ has to be satisfied in order to have $\mathcal{M} > 0$.

    For our purposes, it is convenient to express the internal spacetime coefficients using the parameters of the external spacetime that can be directly determined from observations of accretion and optical phenomena in vicinity of the neutron stars. Since the matching conditions imply the relations
    \begin{equation}\label{EQq}
      q = - \frac{3G M R \varrho}{\lambda} = \frac{3G \mathcal{M}R}{\left(1 - \lambda/\varrho\right)},
    \end{equation}

    \begin{equation}
      \frac{\lambda}{\varrho} = 1-\frac{3G\mathcal{M}R}{q},
    \end{equation}
    the internal metric can be expressed in terms of the external parameters $\mathcal{M}$, $R$, $q$ and a new  parameter
    \begin{equation}\label{EQX}
      X \equiv \frac{q}{3G \mathcal{M}R}
    \end{equation}
    in the form
    \begin{equation}
      \left(B^{-}(r)\right)^2 = \Delta^{-2}(r) = \left[1 - \frac{2G\mathcal{M}r^2}{R^3} \left(1 - \frac{3}{2}X\right)\right]^{-1}
    \end{equation}
    \begin{equation}
      A^{-}(r) = \frac{\Delta(R)\left[3\Delta(R)\left(2X - 1\right)+\Delta(r)\left(1 - 3X\right)\right]}{2\Delta(R)\left(2X - 1\right) - \Delta(r) X}.
    \end{equation}
    The pressure profile can be expressed in the form

    \begin{equation}
      \frac{p(r)}{\varrho} = \frac{\left[\Delta(r) - \Delta(R)\right] \left(2X - 1\right)}{\left[3\Delta(R) - \Delta(r)\right] \left(X - 1\right) + \left[3\Delta(R) - 2\Delta(r)\right] X}.
    \end{equation}
    The observational restrictions obtained from the measurements outside the neutron star then can be applied to the internal parameters using the relations~(\ref{EQq})--(\ref{EQX}).

  \subsection{Limit on existence of uniform density stars}
\begin{figure}[t]
      \centering\includegraphics[width=0.8\hsize,keepaspectratio=true]{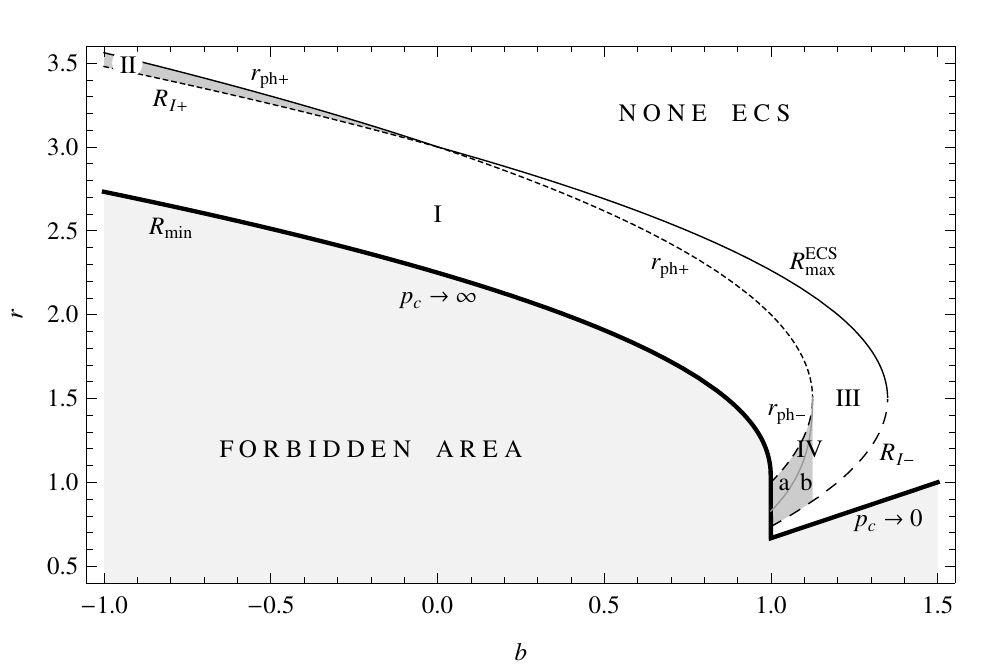}
      \caption{Classification of the extremely compact stars (ECS). The parameter space $(b-R)$ is separated into five Zones~I--IV${}_\mathrm{a, b}$; their properties are given in Section~\ref{SECngaep}.}\label{FIGzone}
    \end{figure}

    The limit on the existence of the uniform density spherical configuration is related to their compactness and is determined by the condition of pressure finiteness in their center. We then find the limit on compactness of the star given in terms of the gravitational mass related to the internal geometry~\cite{Ger-Maar:2001:}:
    \begin{equation}\label{EQGMR}
      \frac{G M}{R}\leq \frac{4}{9}\left[\frac{1 + \frac{5}{4}\frac{\varrho}{\lambda}}{\left(1 + \frac{\varrho}{\lambda}\right)^2}\right].
    \end{equation}
    The lowest order correction is given by
    \begin{equation}
      \frac{G M}{R}\leq \frac{4}{9}\left[1 - \frac{3}{4}\frac{\varrho}{\lambda}\right].
    \end{equation}
    Using the external gravitational mass parameter $\mathcal{M}$, the compactness limit~(\ref{EQGMR}) is transformed to the form
    \begin{equation}
      \frac{G \mathcal{M}}{R}\leq \frac{4}{9}\left[\frac{\left(1 + \frac{5}{4}\frac{\varrho}{\lambda}\right)\left(1 - \frac{\varrho}{\lambda}\right)}{\left(1 + \frac{\varrho}{\lambda}\right)^2}\right].
    \end{equation}
    Introducing the gravitational radius
    \begin{equation}r_\mathrm{g} \equiv G \mathcal{M}\end{equation}
    and dimensionless braneworld tidal charge
    \begin{equation}b \equiv \frac{q}{r_\mathrm{g}^{2}},\end{equation}
    we can put
    \begin{equation}
      X = \frac{q}{3G\mathcal{M}R} = \frac{1}{3}b\frac{r_\mathrm{g}}{R}
    \end{equation}
    and we arrive to the relations
    \begin{equation}
      \Delta^{2}(R) = 1 - 2\frac{r_\mathrm{g}}{R} + b \left(\frac{r_\mathrm{g}}{R}\right)^2\end{equation}
    \begin{equation}
      \Delta^{2}(r) = 1 - \frac{r_\mathrm{g}}{r}\left(\frac{r}{R}\right)^3 \left(2 - b\frac{r_\mathrm{g}}{R}\right)
    \end{equation}
    In terms of dimensionless units ($r_\mathrm{g} = 1$), the pressure function reads
    \begin{equation}
      \frac{p}{\varrho}\left(r, R, b\right) = \frac{\left(\Delta(r) - \Delta(R)\right) \left(2b - 3R\right)}{3 \Delta(R) \left(2b - 3R\right) - 3\Delta(r) \left(b - R\right)}.
    \end{equation}
    The pressure increases monotonously with radius decreasing. The central pressure is given by
    \begin{equation}
      \frac{p}{\varrho}\left(r = 0, R, b\right) = \frac{3R - 2b}{3R \left(\sqrt{b + (R - 2)R} + R - 3\right) + 6b}.
    \end{equation}
    The reality condition of the central pressure reads $b + (R-2) R\geq 0$ and is equivalent to the condition $\Delta(R)\in \mathbb{R}$.
    The pressure in the center of the star must be finite and positive. These conditions imply the limits $R_\mathrm{min}(b)$ on the existence of uniform density stars. The central pressure $p(r = 0, R, b)$ diverges when the surface radius $R$ satisfies the condition
    \begin{equation}
      3R \left(\sqrt{b + (R-2) R} + R-3\right) + 6b = 0.
    \end{equation}
    This leads to a cubic equation relative to $R$ which gives one solution that is relevant for $b\leq 1$ and reads
    \begin{eqnarray}
       R & = & \frac{1}{4} \left(\frac{(b-9) (b-1)}{\sqrt[3]{8 \sqrt{(b-1)^2 b^3}+b \left[b (b+17)-45\right]+27}} + \right. \nonumber \\
        &   & + \left.\sqrt[3]{8 \sqrt{(b-1)^2 b^3}+b \left[b (b+17)-45\right]+27}+b+3\right).
    \end{eqnarray}
    The solution is depicted on Figure~\ref{FIGzone} as the part of $R_\mathrm{min}(b)$ denoted by $p \rightarrow \infty$. For $b = 0$, we arrive at the standard condition $R > \frac{9}{4}$ (see~\cite{Stu-Tor-Hle-Urb:2009:}). For $b > 1$ the relevant limit is given by the condition of positiveness of the central pressure. It reads
    \begin{equation}
      R_\mathrm{min}(b| b > 1) = \frac{2}{3}b
    \end{equation}
    and is denoted by $p \rightarrow 0$. For smaller $R$ the central pressure is not positive.

    Note that for $3/4 < b < 1$ there is a region of surface radii given by the condition $2b/3 < R < 1-\sqrt{1-b}$ where $p(r = 0)$ is positive. However, the star surface is located under the inner horizon of the external spacetime that belongs to the black hole type R-N spacetime. Such a configuration has to be hidden under the inner horizon of the external spacetime and is irrelevant for our considerations.

  \subsection{Extremely compact neutron stars}
    The ECS (neutron, quark, or hybrid) are defined by the existence of trapped null geodesics in the interior of the compact star. In the case of internal uniform density Schwarzschild spacetimes ($b = 0$), such objects appear just when the surface of the compact star is located under the photon circular geodesic of the external vacuum Schwarzschild spacetime located at $r_\mathrm{ph} = 3r_\mathrm{g}$ \cite{Stu-Tor-Hle-Urb:2009:}. In the braneworld uniform density compact stars this simple rule does not hold and the situation is more complex due to different character of the external spacetime and its relation to the internal spacetime.

    Assuming the external spacetime to be of the Reissner-Nordstr\"{o}m (R-N) type with the tidal charge substituting the electric charge squared appearing in the standard R-N spacetimes, the photon circular geodesics are given by the condition
    \begin{equation}
      r^2 - 3r_\mathrm{g}r + 2q = 0.
    \end{equation}

    However, it is convenient to use dimensionless tidal charge $b$ and dimensionless radial coordinate defined by
    \begin{equation}
      r/r_\mathrm{g}\rightarrow r.
    \end{equation}

    For negative tidal charges ($b < 0$) external spacetimes of the black-hole type are allowed only~\cite{Kot-Stu-Tor:2008:CLASQG:}.
    The radii of the photon circular orbits are given by
    \begin{equation}
      r_{\mathrm{ph}\pm} = \frac{3}{2}\left(1 \pm \sqrt{1-\frac{8}{9}b}\right). \label{EQph}
    \end{equation}
    However, only the outer photon circular geodesic at $r_\mathrm{ph+}$ is physically relevant and we see immediately that $(b < 0)$ there is $r_{\mathrm{ph}}>3$.

    For braneworld neutron stars with a positive tidal charge ($b > 0$) both black-hole and naked singularity R-N spacetimes are relevant \cite{Kot-Stu-Tor:2008:CLASQG:}. In the black-hole type spacetimes ($b \leq 1$), the photon circular orbits are given by the relation~(\ref{EQph}), but only the outer solution corresponding to an unstable orbit can be astrophysically relevant since the surface of the compact star has to be located above radius of the outer horizon. In the naked singularity spacetimes ($b > 1$) two photon circular orbits are can be astrophysically relevant when $b < 9/8$, the inner one being stable, the outer one --- unstable. In the spacetimes with $b > 9/8$, no photon circular orbits can exist. Therefore, in the braneworld spacetimes, the existence of ECS is governed by the interplay of the behaviour of the internal and external effective potential of the motion and will be determined and classified in the following section.

  \section{Null geodesics of internal and external spacetimes}\label{SECogeoi}
    In terms of the tetrad formalism the metric~(\ref{EQmetric}) reads
    \begin{equation}
      \mathrm{d} s^{2} = -[\omega^{(t)}]^{2}
                         +[\omega^{(r)}]^{2}
                         +[\omega^{(\theta)}]^{2}
                         +[\omega^{(\phi)}]^{2},
    \end{equation}
    where
    \begin{equation}
      \omega^{(t)}  = A(r)\,\mathrm{d} t,\quad
      \omega^{(r)}      = B(r)\,\mathrm{d} r,\quad
      \omega^{(\theta)} = r\,\mathrm{d} \theta,\quad
      \omega^{(\phi)}   = r\sin\theta\,\mathrm{d} \phi.
    \end{equation}
    Since there is $e^{\mu}_{(\alpha)} = \left[\omega^{(\alpha)}_{\mu}\right]^{-1}$,
    the tetrad of 4-vectors is given by
    \begin{equation}
      \vec{e}_{(t)}  = \frac{1}{A(r)} \frac{\partial}{\partial t},\quad
      \vec{e}_{(r)}      = \frac{1}{B(r)} \frac{\partial}{\partial r},\quad
      \vec{e}_{(\theta)} = \frac{1}{r}    \frac{\partial}{\partial \theta},\quad
      \vec{e}_{(\phi)}   = \frac{1}{r\sin\theta}\frac{\partial}{\partial \phi}.
    \end{equation}
    Tetrad components of 4-momentum of a test particle or a photon are determined by the projections $p_{(\alpha)} = p_{\mu}e^{\mu}_{(\alpha)}$, $p^{(\alpha)} = p^{\mu}\omega_{\mu}^{(\alpha)}$ which give quantities measured by the local observers.

  \subsection{Null geodesics and effective potential}\label{SECngaep}
    We consider the period of evolution and cooling of ECS when their temperature falls down enough that the motion of neutrinos can be considered free, i.e., geodetical. We can assume this period starts at the moment when mean free path of neutrinos becomes to be comparable to the radius $R$, i.e., in hours after the gravitational collapse creating the compact object \cite{Gle:1992:PHYSR4:,Gle:2000:CompactStars:,Hae-Zdu:1986:NATURE:,Web:1999:Pul:,Sha-Teu:1983:BHWDNS:}. In fact, there are arguments that this condition starts to be fulfilled about 50~s after collapse to a proto-neutron star \cite{Lat-Pra:2007:PhysRep:,Lat-Pra:2004:SCIENCE:}. Weak interaction of ultrarelativistic (massless) neutrinos implies their motion along null geodesics obeying the equations ($\lambda$ is an affine parameter)
    \begin{equation}
      \frac{\mathrm{D} p^{\mu}}{\mathrm{d} \lambda} = 0,\qquad
      p^{\mu}p_{\mu} = 0.                                          \label{EQgeod}
    \end{equation}

    Due to the existence of two Killing vector fields: the temporal $\partial/\partial t$ one, and the azimuthal $\partial/\partial \phi$ one, two conserved components of the 4-momentum must exist:
    \begin{equation}
      E=-p_{t}\quad\mbox{(energy),}\qquad
      L=p_{\phi}\quad\mbox{(axial angular momentum)}.              \label{EQconserv}
    \end{equation}
    Moreover, the motion plane is central. For a single-particle motion, one can set $\theta=\pi/2=\mathrm{const}$, choosing the equatorial plane.

    The motion along null-geodesics is independent of energy (frequency) and can conveniently be described in terms of the impact parameter
    \begin{equation}
      \ell = \frac{L}{E}.                                           \label{EQimpar}
    \end{equation}
    Then~(\ref{EQgeod}) yields the relevant equation governing the radial motion in the form
    \begin{equation}
      (p^{r})^{2} = A^{-2}(r)B^{-2}(r)E^{2}
        \left(1-A^{2}(r)\frac{\ell^{2}}{r^{2}}\right).              \label{EQgovmot}
    \end{equation}
    Clearly, the energy $E$ is irrelevant and can be used for rescalling of the affine parameter $\lambda$. The radial motion is restricted by an effective potential related to the impact parameter $\ell$, and defined by the relations
    \begin{equation}
      \ell^{2} \leq V{}_{\mathrm{eff}} =\left\{
      \begin{array}{lll}
        V{}_{\mathrm{eff}}^{\mathrm{int}} = \displaystyle\frac{r^2}{(A^{-}(r))^2}  & \quad\mbox{for} & r\leq R\\
        V{}_{\mathrm{eff}}^{\mathrm{ext}} = \displaystyle\frac{r^2}{(A^{+}(r))^2}  & \quad\mbox{for} & r > R.
      \end{array}\right.
    \end{equation}
    $V{}_{\mathrm{eff}}^{\mathrm{int}}$ is the effective potential of the null-geodetical motion in the internal  braneworld spacetime and $V{}_{\mathrm{eff}}^{\mathrm{ext}}$ is the effective potential of the null-geodetical motion in the external, vacuum R-N spacetime. We shall see in the following that due to the bulk space tidal effects on the matching of the internal and external spacetimes (see \cite{Ger-Maar:2001:} for details) we obtain a non-standard variety of relations of the effective potentials in the ECS interior and exterior.

    Using the dimensionless radial coordinate expressed in terms of the gravitational radius ($r/G\mathcal{M}\rightarrow r$), and dimensionless tidal charge $b$, we obtain the relations

    \begin{equation}
      V{}_{\mathrm{eff}}^{\mathrm{int}} = \frac{r^2 R^2 \left[b Y - 2R (2b - 3R) Z\right]^2}{9 \left[b + (R-2) R\right] \left[(R - b) + R (2b -3R) Z\right]^2}, \label{EQellsb}
    \end{equation}
    \begin{equation}
      V{}_{\mathrm{eff}}^{\mathrm{ext}} = \frac{r^4}{b + (r - 2) r},
    \end{equation}
    where
    \begin{eqnarray}
      Y &\equiv & R^2 \Delta(r) = \sqrt{r^2(b-2R)+R^4},\\
      Z &\equiv & R \Delta(R) = \sqrt{b+(R-2) R}.
    \end{eqnarray}

    \begin{figure}[t]%
      \begin{minipage}[b]{.499\hsize}
        \centering\includegraphics[width=\hsize,keepaspectratio=true]{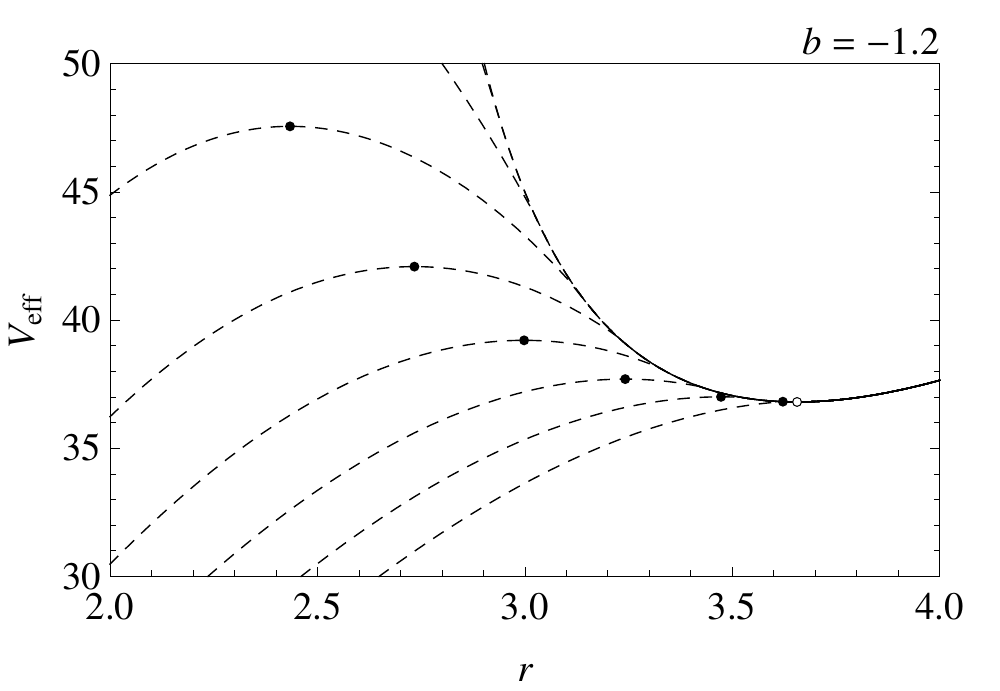}
        \centering\includegraphics[width=\hsize,keepaspectratio=true]{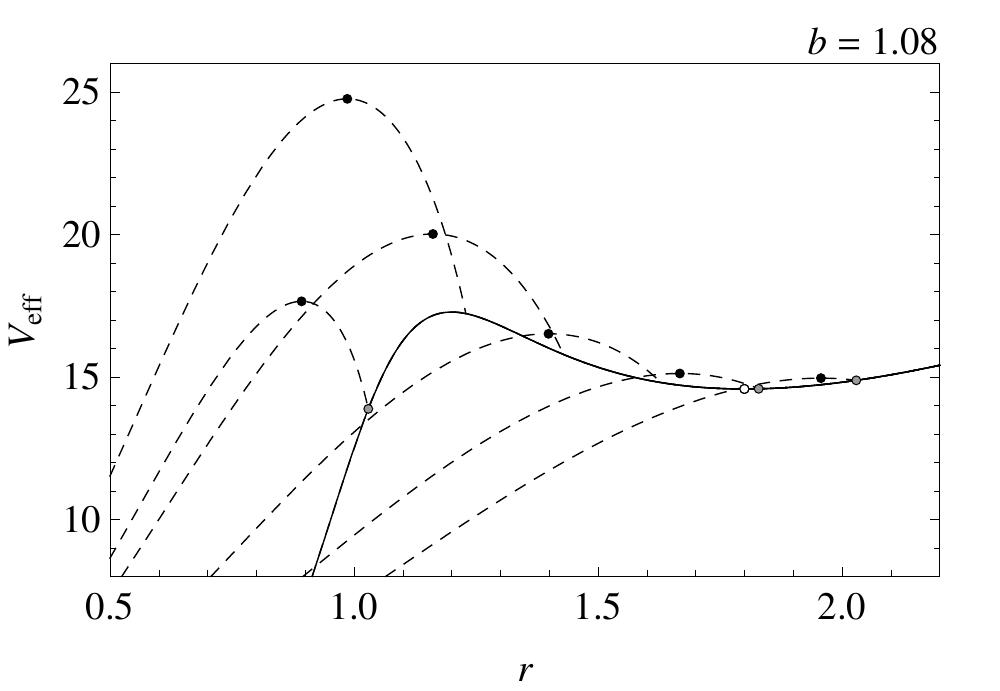}
      \end{minipage}\hfill%
      \begin{minipage}[b]{.499\hsize}
        \centering\includegraphics[width=\hsize,keepaspectratio=true]{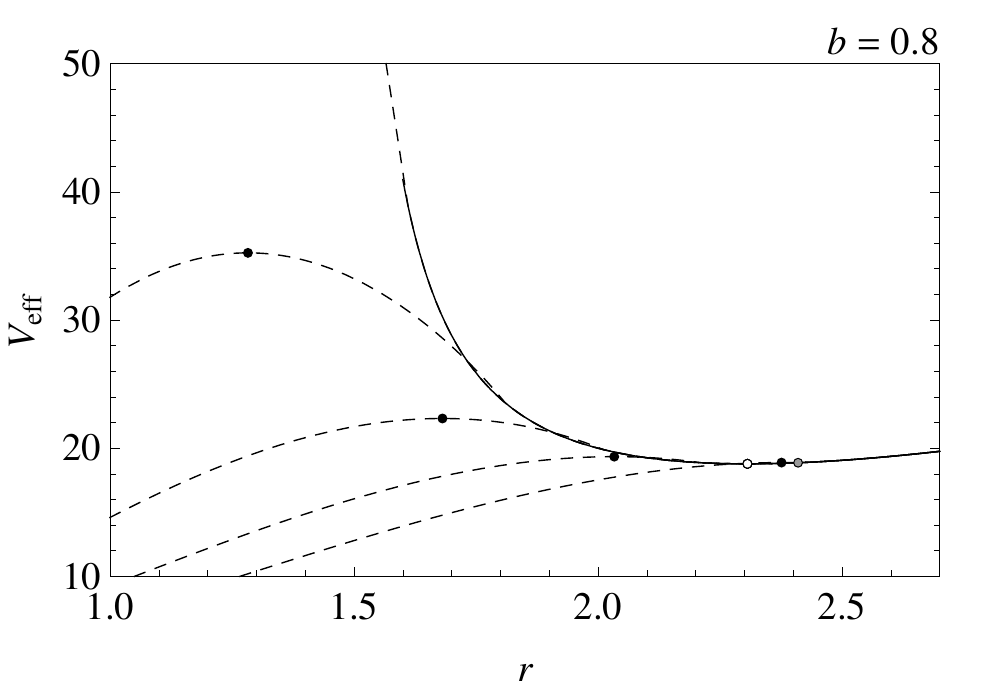}
        \centering\includegraphics[width=\hsize,keepaspectratio=true]{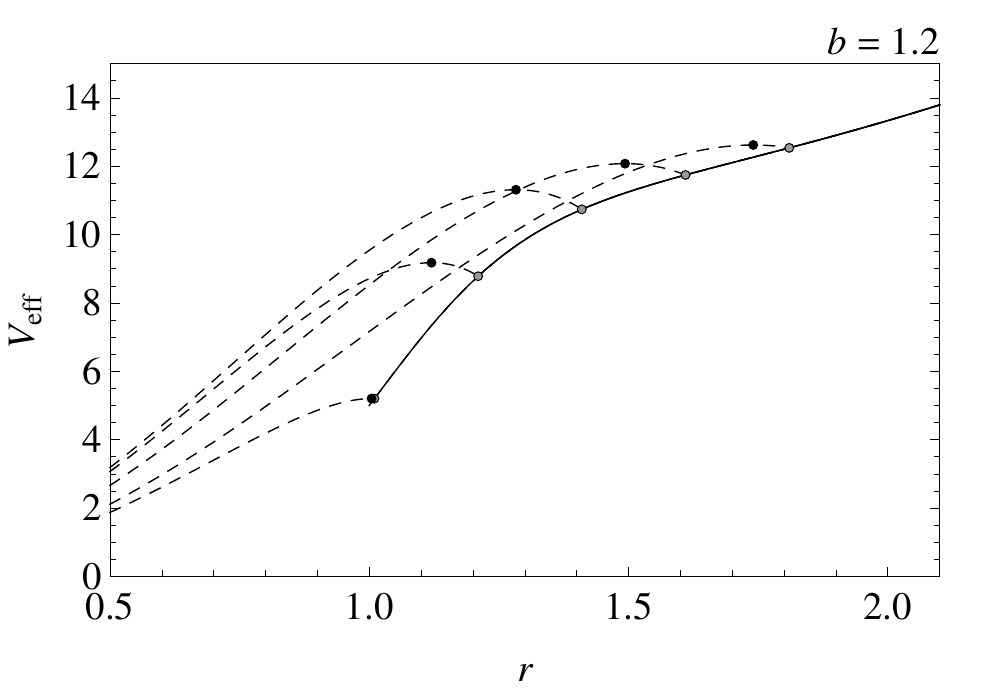}
      \end{minipage}
      \centering\includegraphics[width=.49\hsize,keepaspectratio=true]{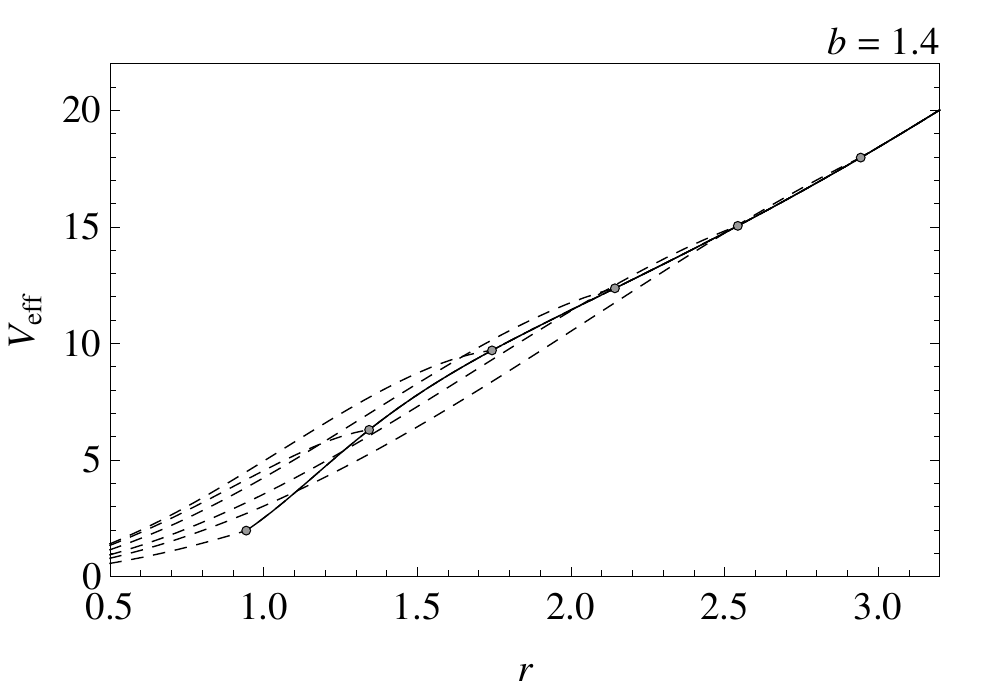}
      \caption{The effective potential $V_{\mathrm{eff}}$ of both internal (dashed lines) and external spacetimes (full lines) given for the black-hole type ($b < 1$) and naked-singularity ($b > 1$) type of the external spacetime. All characteristic cases of its behaviour are presented for both ECS and non-extreme compact stars.}\label{FIGVdiff}
    \end{figure}

    Circular null geodesics, located at $r_\mathrm{c(i)}$ and $r_\mathrm{c(e)}$, respectively, are given by the local extrema of the effective potential ($\partial {V_{\mathrm{eff}}}/\partial {r} = 0$). In the internal spacetime we have to solve a nontrivial equation $r$:
    \begin{eqnarray}
      \lefteqn{\mathrm{d}\left(V{}_\mathrm{eff}^\mathrm{int}(r, R, b)\right)/\mathrm{d}r  \equiv}\nonumber \\
      & & \left[2 r R^2 \left(b (Y - 4 R Z) + 6 R^2 Z\right) \left\{-b^3 \left(r^2 (Y-4 R Z)+8 R^2 Y\right) + b^2 R\times \right.\right. \nonumber \\
      & & \times \left(r^2 (3 Y-14 R Z)+R^2 \left(6 R^2 Z - 9 R Y + 40 Y\right)\right)+b R^2 \left(R^2(-13 R^2 Z + \right.\nonumber \\
      & & \left.\left.\left.\left. + 25 R Y- 66 Y\right)- 2 r^2 (Y- 6 R Z)\right)+ 6 R^5 \left(R^2 Z-3 R Y+ 6 Y\right)\right\}\right]\times \nonumber \\
      & &\times \left[9 \left(b+(R-2) R\right) Y \left(R (Y-3 R Z)-b \left(Y-\nonumber 2 R Z\right)\right)^3\right]^{-1} = 0.\\
    \end{eqnarray}
    This equation can be solved by using numerical methods and determines the loci $r_\mathrm{c(i)}(R, b)$ of the internal circular null geodesics that are stable and correspond to a local maximum of the internal effective potential. From the radius $r_\mathrm{c(i)}$ we easily obtain the impact parameter $\ell^{2}_{\mathrm{c(i)}}(R, b)$ using equation~(\ref{EQellsb}).  The existence of the internal stable circular null geodesics is allowed only for the surface radius $R$ limited by values explicitly given by
    \begin{equation}
       R_\mathrm{I\pm} = \frac{3}{2} \pm \sqrt{\frac{9}{4} - \frac{5}{3}b}
    \end{equation}
    that has to satisfy naturally also the condition $R > R_\mathrm{min}(b)$.

    In the external spacetime, the photon circular geodesics and the related local extrema of the effective potential are given by equation~(\ref{EQph}) and their properties were discussed in the previous section.

    When we take into account the (non-)existence of the local extrema of the internal/external part of the effective potential $V_\mathrm{eff}$, we obtain the whole region of existence of ECS.
    The function $R{}_\mathrm{max}^\mathrm{ECS}(b)$ is determined by the relation
    \begin{equation}
      R{}_\mathrm{max}^{\mathrm{ECS}} =\left\{
        \begin{array}{lll}
          r_\mathrm{ph} \equiv \frac{1}{2} \left(3 + \sqrt{9-8 b}\right) & \quad\mbox{for} & b \leq 0\\
          R_\mathrm{I+} \equiv \frac{3}{2} + \sqrt{\frac{9}{4} - \frac{5}{3}b}  & \quad\mbox{for} & 0 < b < 27/20
        \end{array}
      \right.,
    \end{equation}
    while for $R{}_\mathrm{min}^\mathrm{ECS}(b)$ we have
    \begin{equation}
      R{}_\mathrm{min}^{\mathrm{ECS}} = \left\{
        \begin{array}{lll}
          R_\mathrm{min} & \quad\mbox{for} & b \leq 1\\
          R_\mathrm{I-} \equiv \frac{3}{2} - \sqrt{\frac{9}{4} - \frac{5}{3}b} & \quad\mbox{for}& 1 < b < 27/20
        \end{array}
      \right..
    \end{equation}
    The region of ECS in the parameter space is represented in Figure~\ref{FIGzone}.

    Typical behaviour of the effective potential of the null-geodetical motion $V{}_{\mathrm{eff}}$ is demonstrated in Figure~\ref{FIGVdiff} for appropriatelly chosen values of the tidal charge $b$. Here, and henceforth, we express the radii in units of the gravitational radius $r_g = G\mathcal{M}$. The selection of the tidal charge values used in Figure~\ref{FIGVdiff} demonstrates the full classification of the behavior of the effective potential in both internal and external spacetimes. When the effective potential in the internal or external spacetime (or in both of them) has a local extreme corresponding to a photon circular geodesic, trapped null geodesics can appear if the surface radius $R$ is properly chosen giving an ECS. On the other hand, we find an ordinary compact star if the surface is chosen in such a way that no local extrema of the effective potentials exist (see Figure~\ref{FIGVdiff}e). Separation of the zones of compact stars of different character, both extremely compact and ordinary compact, in the parameter space $(b-R)$, is determined by the functions $R_\mathrm{min}(b)$, $R_\mathrm{max}(b)$ and $r_\mathrm{ph}(b)$ where the last function governs radius of the photon circular geodesics in the external spacetime (both of them for $1 < b < 9/8$). The subdivision of the Zone~IV in the tidal charge range $1 < b < 9/8$ is given by the relation of the magnitude of the effective potential at the surface of the ECS and the magnitude of the external effective potential at its local minimum, and is determined numerically (see Figure~\ref{FIGzone}).

    We divided the region of the parameter space $R-b$ corresponding to the existence of ECS into five zones. The classification is based on the existence of local maxima/minima of $V{}^{\mathrm{int/ext}}_{\mathrm{eff}}$ in the following way (see Figures~\ref{FIGzone} and \ref{FIGbehzoneALL})
    \begin{itemize}
      \item[$\bullet$]{\emph{Zone I} there exist maximum of $V{}^{\mathrm{int}}_{\mathrm{eff}}$ and minimum of $V{}^{\mathrm{ext}}_{\mathrm{eff}}$, both located at $r \neq R$, $\mathrm{min} V{}^{\mathrm{ext}}_{\mathrm{eff}} < V{}^{\mathrm{ext}}_{\mathrm{eff}}(r = R)$, and there is no local maximum of $V{}^{\mathrm{ext}}_{\mathrm{eff}}$ at $r > R$;}
      \item[$\bullet$]{\emph{Zone II} there exist maximum of $V{}^{\mathrm{int}}_{\mathrm{eff}}$ at $r = R$  and minimum of $V{}^{\mathrm{ext}}_{\mathrm{eff}}$ at $r \neq R$, $\mathrm{min} V{}^{\mathrm{ext}}_{\mathrm{eff}} < \mathrm{max} V{}^{\mathrm{int}}_{\mathrm{eff}}$, and there is no local maximum of $V{}^{\mathrm{ext}}_{\mathrm{eff}}$ at $r > R$;}
      \item[$\bullet$]{\emph{Zone III} there exist maximum of $V{}^{\mathrm{int}}_{\mathrm{eff}}$ at $r < R$  and minimum of $V{}^{\mathrm{ext}}_{\mathrm{eff}}$ at $r = R$; there is no local maximum of $V{}^{\mathrm{ext}}_{\mathrm{eff}}$ at $r > R$;}
      \item[$\bullet$]{\emph{Zone IV} there exist both the local maximum and local minimum of $V{}^{\mathrm{ext}}_{\mathrm{eff}}$ at $r > R$, this zone can be divided into two parts
          \begin{itemize}
            \item[a)]{minimum of $V{}^{\mathrm{ext}}_{\mathrm{eff}} < V{}^{\mathrm{int}}_{\mathrm{eff}}\left(r = R\right)$;}
            \item[b)]{ minimum of $V{}^{\mathrm{ext}}_{\mathrm{eff}} > V{}^{\mathrm{int}}_{\mathrm{eff}}\left(r = R\right)$.}
          \end{itemize}}
    \end{itemize}

  \section{Trapping of neutrinos}\label{SECtrappn}
\begin{figure}[t]%%%%%%%%%%%%%%%%%%%%%%%%%%%%%%%%%%%%%%%
      \begin{minipage}[b]{.499\hsize}
        \centering\includegraphics[width=\hsize,keepaspectratio=true]{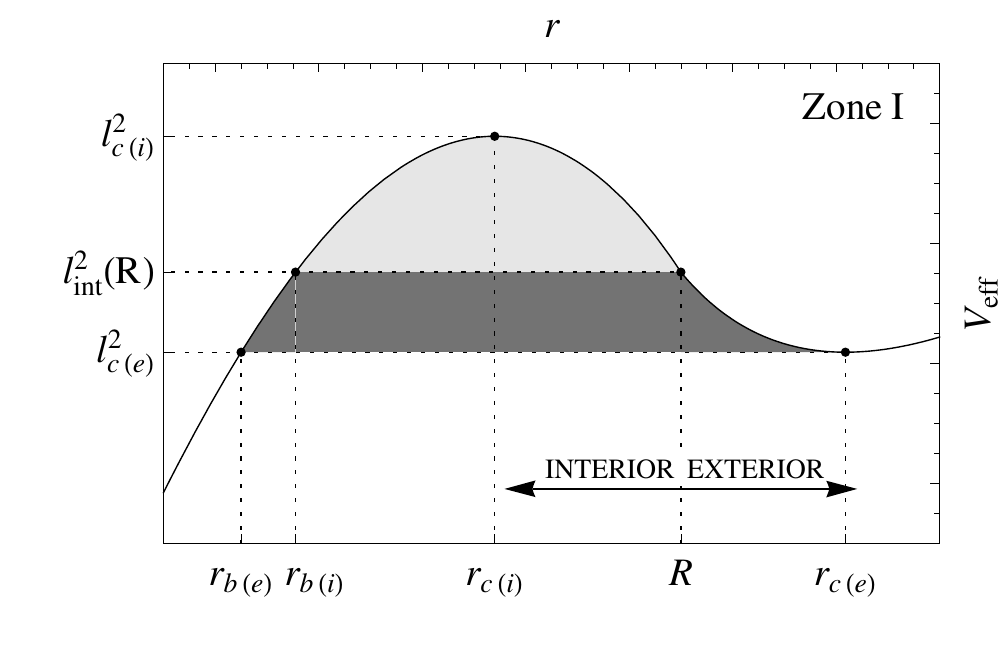}
        \centering\includegraphics[width=\hsize,keepaspectratio=true]{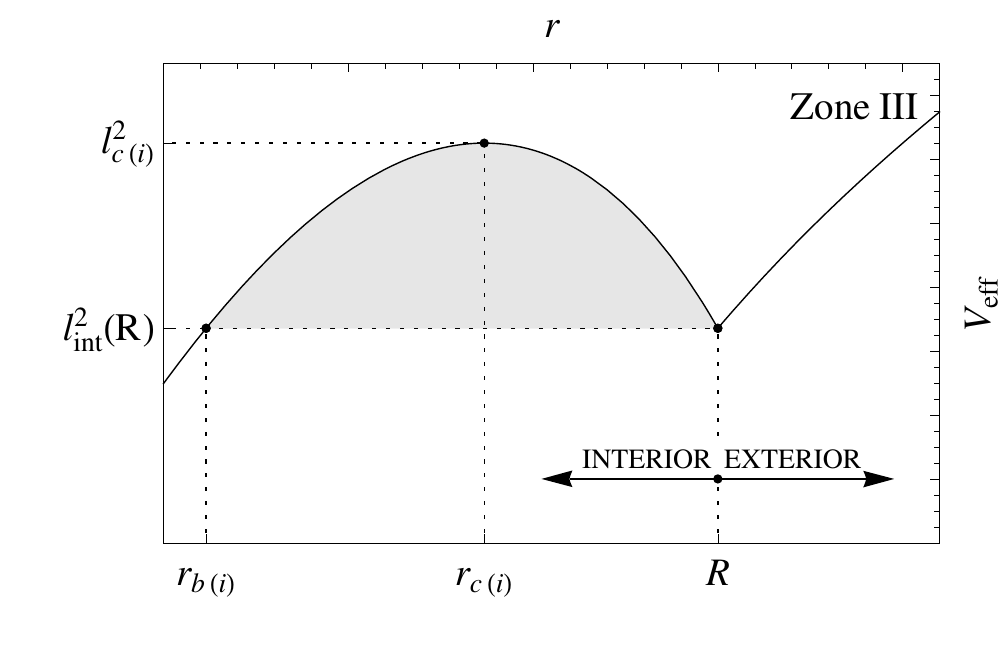}
      \end{minipage}\hfill%
      \begin{minipage}[b]{.499\hsize}
        \centering\includegraphics[width=\hsize,keepaspectratio=true]{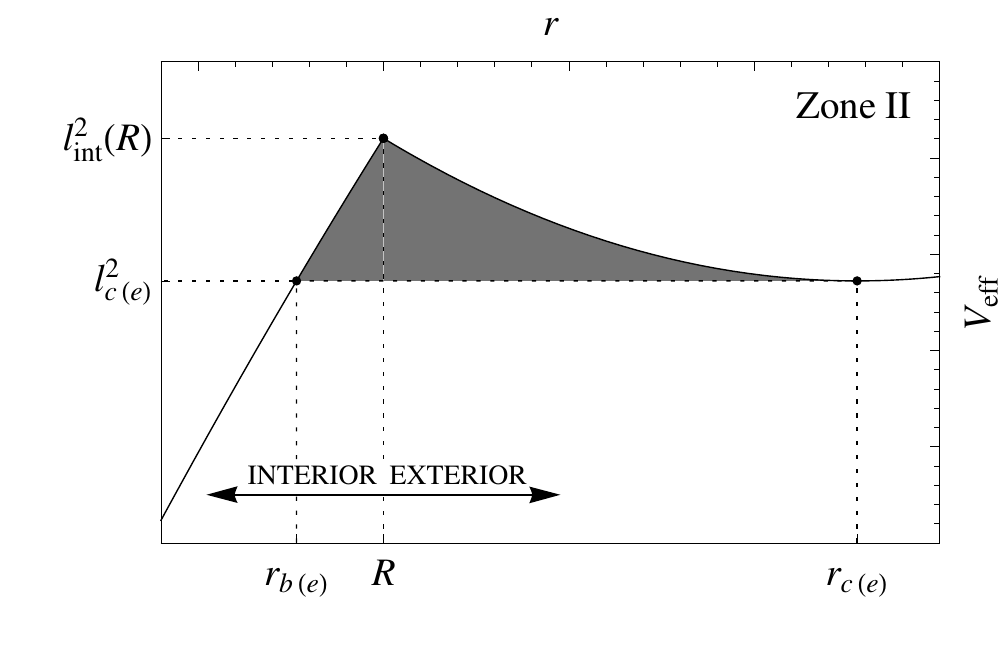}
        \centering\includegraphics[width=\hsize,keepaspectratio=true]{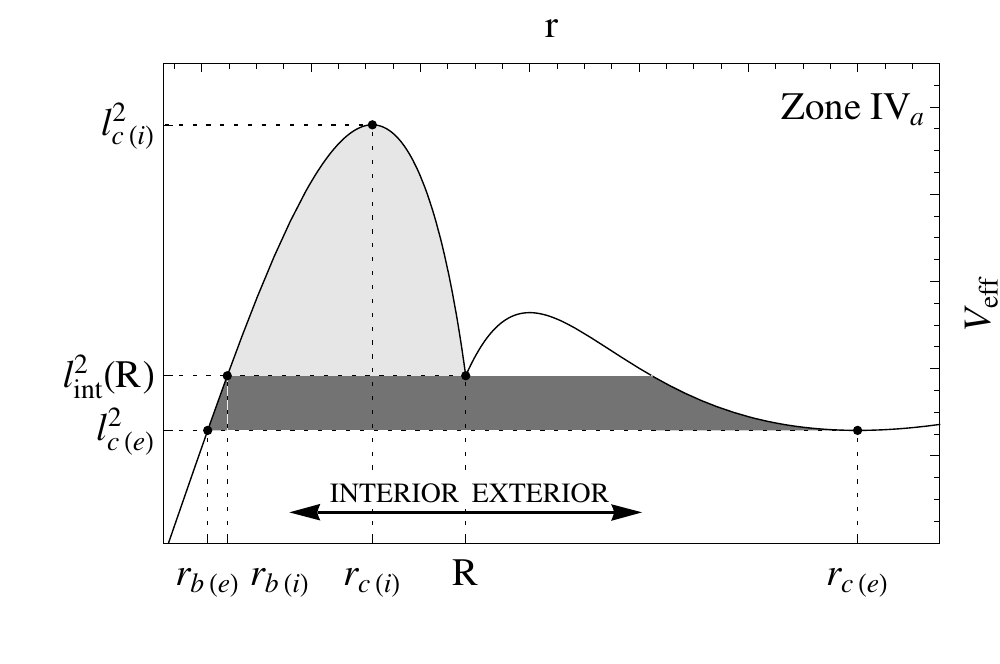}
      \end{minipage}
      \centering\includegraphics[width=0.499\hsize,keepaspectratio=true]{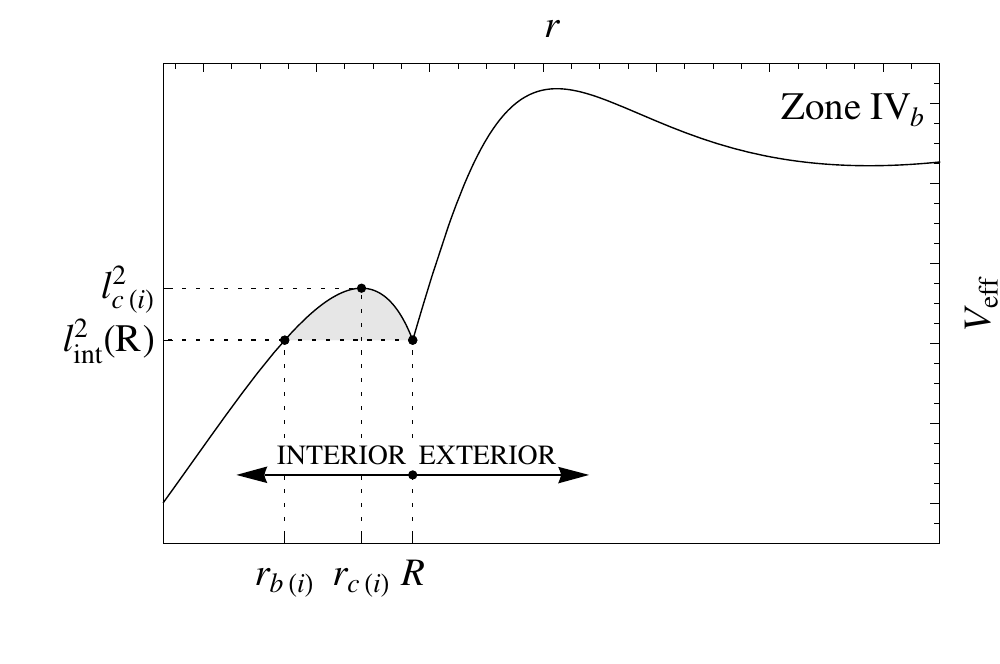}
      \caption{Schematic classification of the effective potential $V{}_\mathrm{eff}$ of the ECS for all the Zones~I--IV$_\mathrm{a, b}$ of the parameter space $(b-R)$ as depicted on Figure~\ref{FIGzone}.}\label{FIGbehzoneALL}
    \end{figure}

    In ECS, some part of produced neutrinos is prevented from escaping these static objects because of stable circular null geodesics existing in the internal spacetime, and/or because of unstable circular null geodesics existing in the external spacetime (Figure~\ref{FIGbehzoneALL}). For external braneworld naked-singularity spacetimes with tidal charge in the interval $1 < b < 9/8$, an additional stable circular null geodesic exists that can be important for appearance and structure of accretion discs (\cite{Stu-Schee:2010:CLAQG:}), but it is irrelevant for trapping of neutrinos radiated by ECS.

    The relation $\partial {V{}^{int}_{\mathrm{eff}}}/\partial {r} = 0$ determines location of the stable circular null geo\-de\-sics of the internal spacetime and implies the impact parameter which corresponds to the local maximum of the effective potential $V{}_{\mathrm{eff}}^{\mathrm{int}}$ at $r_{\mathrm{c(i)}}$ to be given by $\ell_{\mathrm{c(i)}}^{2}$.

    For $b \leq 9/8$, the local minimum of $V{}_{\mathrm{eff}}^{\mathrm{ext}}$ is located at
    \begin{equation}
      r_{\mathrm{c(e)}}=\frac{1}{2} \left(3 + \sqrt{9 - 8 b}\right)
    \end{equation}
    and corresponds to the unstable circular null geodesics of the Reissner-Nord\-str\"{o}m external spacetime, with the impact parameter determined by
    \begin{equation}
      \ell_{\mathrm{c(e)}}^{2} = \frac{\left(3 + \sqrt{9 - 8 b}\right)^4}{8 \left(3 - 2 b + \sqrt{9 - 8 b}\right)}.
    \end{equation}
    For the ECS spacetimes corresponding to all of the Zones~I--IV${}_\mathrm{a, b}$ of the parameter space $(b-R)$, typical behaviour of the effective potentials $V{}_{\mathrm{eff}}^{\mathrm{int}}$ and $V{}_{\mathrm{eff}}^{\mathrm{ext}}$ is represented in Figure~\ref{FIGbehzoneALL} where the regions of trapped neutrinos are shown explicitly.

  \subsection{Regions of trapping}
    Bound neutrinos (depicted by the gray filled areas in Figure~\ref{FIGbehzoneALL}) may be fully trapped inside the internal spacetime, or may, in some spacetimes, appear outside the interior of the ECS being trapped by its strong gravitational field, and entering the interior again; note that the trapped neutrinos moving outside the compact star occur in the spacetimes of the Zones~I, II and IV${}_\mathrm{a}$. Therefore, we divide the trapped neutrinos into two families:
    \begin{itemize}
      \item[$\bullet$] \emph{``Internal'' bound neutrinos} (depicted  using light gray filled area) with
        impact parameter between $\ell^{2}_{\mathrm{int}}(R)$ and
        $\ell^{2}_{\mathrm{c(i)}}$; motion of these neutrinos is restricted to the interior of the ECS. The internal bound neutrinos appear exclusively in spacetimes of the Zones~III and IV${}_\mathrm{b}$, but they occur also in the spacetimes of the Zones~I and IV${}_\mathrm{a}$.
      \item[$\bullet$] \emph{``External'' bound neutrinos} (depicted by the dark gray filling) with impact
        parameter between $\ell^{2}_{\mathrm{c(e)}}$ and
        $\ell^{2}_{\mathrm{int}}(R)$; such bound neutrinos may leave the ECS interior, but they re-enter it. In spacetimes of the Zone~II all bound neutrinos are of this type.
    \end{itemize}

    \begin{figure}[t]%%%%%%%%%%%%%%%%%%%%%%%%%%%%%%%%%%%%%%%
      \begin{minipage}[b]{.499\hsize}
        \centering\includegraphics[width=\hsize,keepaspectratio=true]{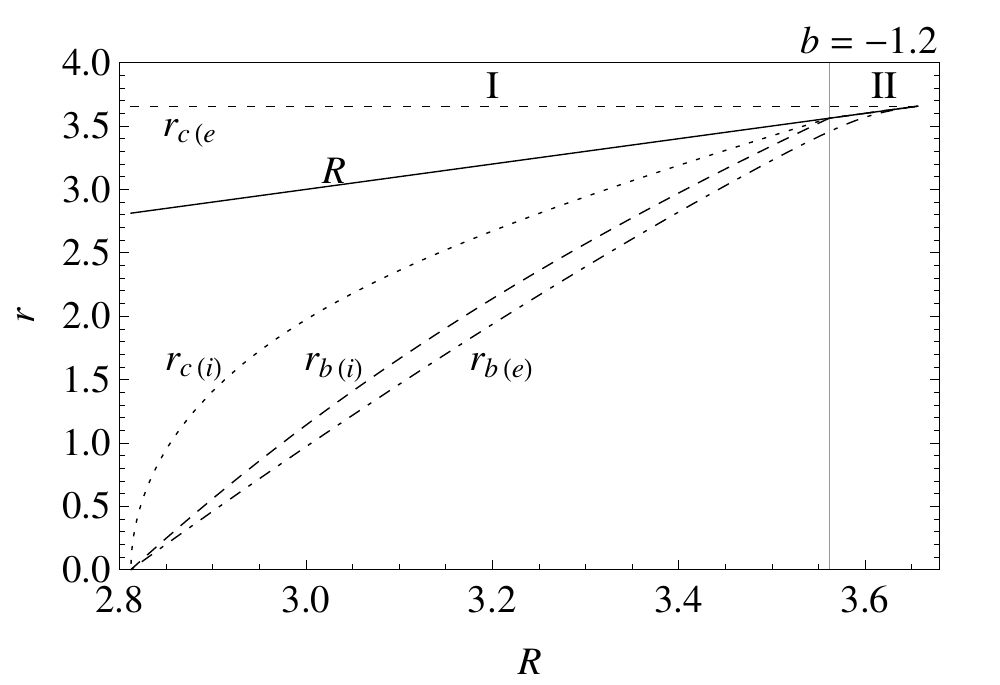}
        \centering\includegraphics[width=\hsize,keepaspectratio=true]{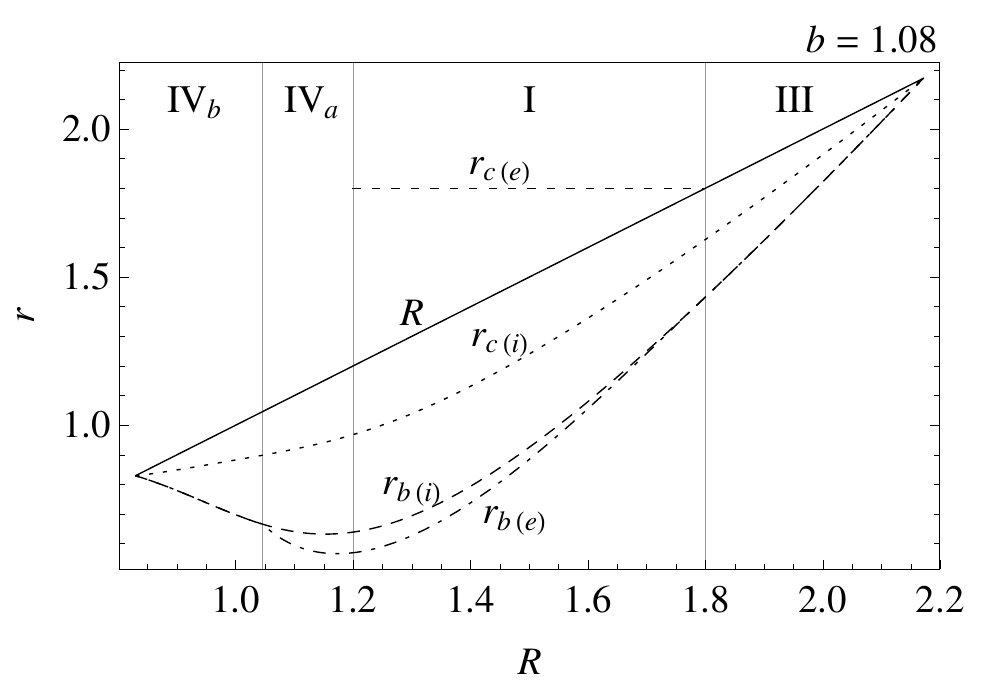}
      \end{minipage}\hfill%
      \begin{minipage}[b]{.499\hsize}
        \centering\includegraphics[width=\hsize,keepaspectratio=true]{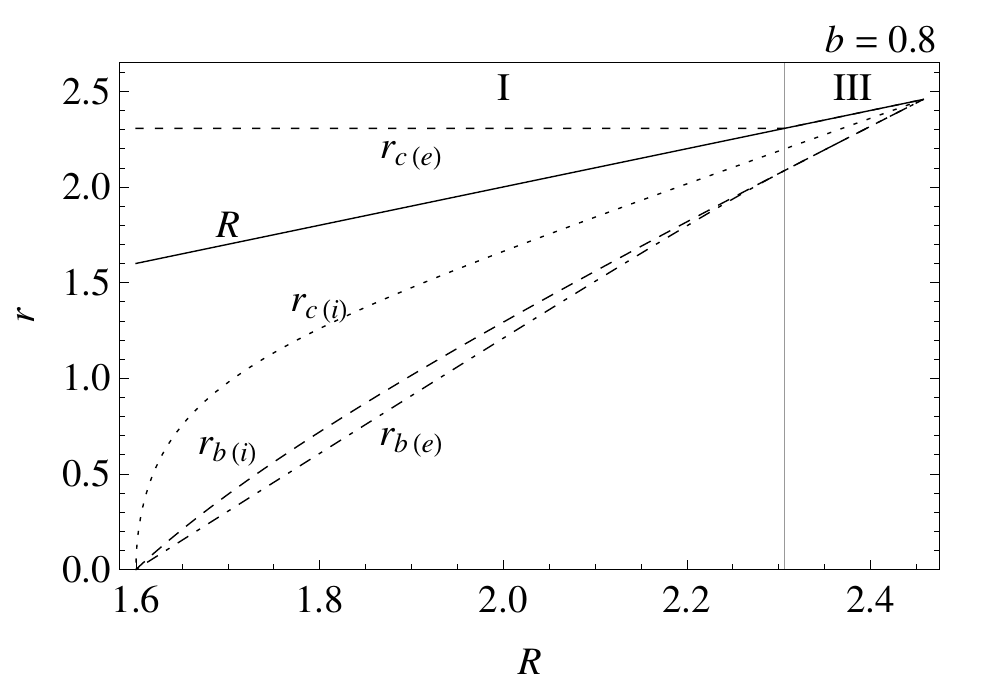}
        \centering\includegraphics[width=\hsize,keepaspectratio=true]{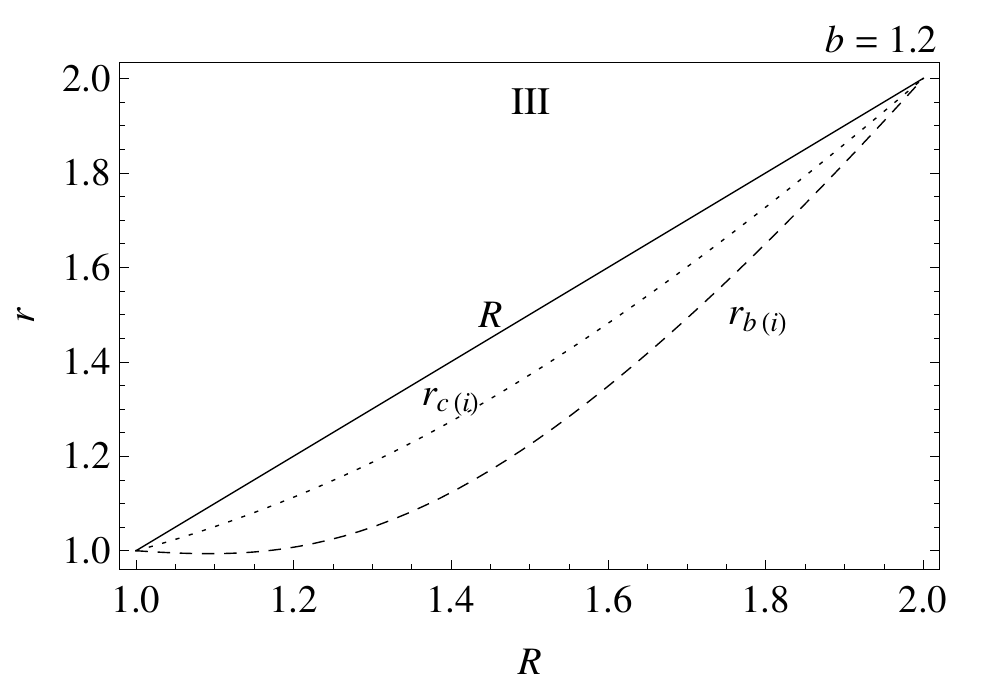}
      \end{minipage}
      \caption{Location of the radii relevant for the trapping of neutrinos $r_\mathrm{b(i)}$, $r_\mathrm{b(e)}$ and  $r_\mathrm{c(e)}$ given as functions of the surface radius $R$ for characteristic values of the tidal charge $b$. The range of the radius $R$ corresponding to the Zones~I--IV of the parameter space $(b-R)$ is depicted in all presented figures.}\label{FIGradius}
    \end{figure}

    The pericentre and the impact parameter of the ``external'' marginally bound neutrinos ($r_{\mathrm{b(e)}}$; Zones~I, II, IV${}_\mathrm{a}$) can be determined from the condition
    \begin{equation}
      V{}^{\mathrm{int}}_{\mathrm{eff}}(r = r_{\mathrm{b(e)}}, R, b) = \frac{\left(3 + \sqrt{9-8 b}\right)^4}{8 \left(3 - 2 b+\sqrt{9-8 b}\right)}\label{EQrbe}
    \end{equation}
    while the pericentre and the impact parameter of the ``internal'' marginally bound neutrinos ($r_{\mathrm{b(i)}}$; Zones~I, III, IV${}_\mathrm{a}$, IV${}_\mathrm{b}$) are given by the condition
    \begin{equation}
      V{}^{\mathrm{int}}_{\mathrm{eff}}(r = r_{\mathrm{b(i)}}, R, b) = \frac{R^4}{b + (R - 2) R},\label{EQrbi}
    \end{equation}
    see Figures~\ref{FIGbehzoneALL},~\ref{FIGradius} for the graphical representation. For completeness, we show in Fig.~\ref{FIGradius} also loci  $r_\mathrm{c(i)}$ of the stable circular null geodesic and the surface radius $R$ of the ECS.

  \subsection{Directional angles}
    Considering (without loss of generality, as stated above equation~(\ref{EQimpar})) an equatorial motion, we can define the \emph{directional angle} relative to the outward pointed radial direction measured in the emitter system (i.e., the local system of static observers in the internal spacetime) by the standard relations
    \begin{equation}
      \sin\psi = \frac{p^{(\phi)}}{p^{(t)}}, \qquad
      \cos\psi = \frac{p^{(r)}}{p^{(t)}},
    \end{equation}
    where
    \begin{equation}
      p^{(\mathrm{\alpha})} = p^{\mu}\omega_{\mu}^{(\mathrm{\alpha})},\qquad
      p_{(\mathrm{\alpha})} = p_{\mu}e^{\mu}_{(\mathrm{\alpha})}
    \end{equation}
    are the neutrino momentum component as measured by the static observers. Besides conserving components~(\ref{EQconserv}), and $p_{\theta} = 0$, equation~(\ref{EQgovmot}) implies
    \begin{equation}
      p_{r} = \pm E A^{-1}B\left(1-A^2 \frac{\ell^{2}}{r^{2}}\right)^{1/2}.
    \end{equation}
    For the directional angles we thus obtain relations
    \begin{equation}
      \sin\psi = A\frac{\ell}{r},\qquad \cos\psi = \pm\left(1-\sin^{2}\psi\right)^{1/2},\label{EQcossin}
    \end{equation}
    where $A$ is given by equation~(\ref{EQamin}).

    The directional angle limit for the bound neutrinos is determined for spacetimes in the Zones~I, II and IV${}_\mathrm{a}$ by the impact parameter $\ell_{\mathrm{c(e)}}^{2}$. We arrive to the relation
    \begin{eqnarray}
       \cos\psi_{\mathrm{e}} = \pm \frac{1}{2} \sqrt{4-\frac{9 \left(\sqrt{9-8 b}+3\right)^4 Z^2 \left[(R-b) Y+R (2 b-3 R) Z\right]^2}{2 \left(-2 b+\sqrt{9-8 b}+3\right) r^2 R^2 \left[b Y+2 R (3 R-2 b) Z\right]^2}}.\label{EQsicose}
    \end{eqnarray}
    The interval of relevant radii is given by $r \in (r_{\mathrm{b(e)}}, R)$.

    The directional angle limit for the ``internal'' bound neutrinos is determined by equation~(\ref{EQcossin}), where for $\ell$ we use $\ell = \ell_\mathrm{int}(R)$. In the Zone~II  spacetimes, there are none neutrinos bound only in their interior. On the other hand, for spacetimes in the Zones~III and IV${}_\mathrm{b}$, this limit represents the total limit on all bound neutrinos. For the internal neutrinos we arrive to the relation
    \begin{eqnarray}
    \cos\psi_{\mathrm{i}} = \pm \sqrt{1-\frac{9 R^2 \left[(R-b) Y + R (2 b-3 R) Z\right]^2}{r^2 \left[b Y + 2 R (3 R-2 b) Z\right]^2}}.
    \end{eqnarray}
    Apparently, the condition $\psi_\mathrm{i} > \psi_\mathrm{e}$ holds at any given radius $r < R$ of the ECS spacetimes belonging to the Zones of the parameter space where both values have good meaning.

  \subsection{Local escaped to produced neutrinos ratio}
    \begin{figure}[t]
    \centering\includegraphics[width=.45\hsize,keepaspectratio=true]{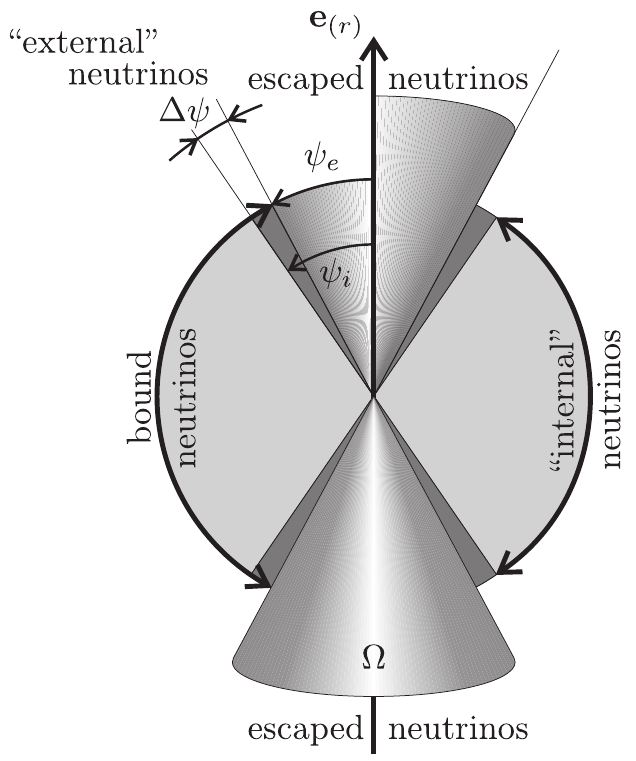}
    \caption{Schematic illustration of the bound-escape ratio at a radius $r \in (r_{\mathrm{b(e)}}, R)$ of an internal spacetime. Direction of the neutrino motion with respect to the static observers is related to $e_{(r)}$ giving the outward oriented radial direction. In the spacetimes of the Zones~III and IV${}_\mathrm{b}$ only $\psi_\mathrm{i}$ cone is relevant, while for spacetime of the Zone~II only angle $\psi_\mathrm{e}$ is relevant.}\label{FIGbound}
    \end{figure}

    We assume that neutrinos are locally produced by isotropically emitting sources. Then escaped-to-produced-neutrinos ratio depends on a geometrical argument only. It is determined by the solid angle $2\Omega$ corresponding to escaping neutrinos (also inward emitted neutrinos must be involved because even these neutrinos can be radiated away), see Figure~\ref{FIGbound}.

    \begin{figure}[ht]
      \begin{center}
         \includegraphics[width=.95\hsize,keepaspectratio=true]{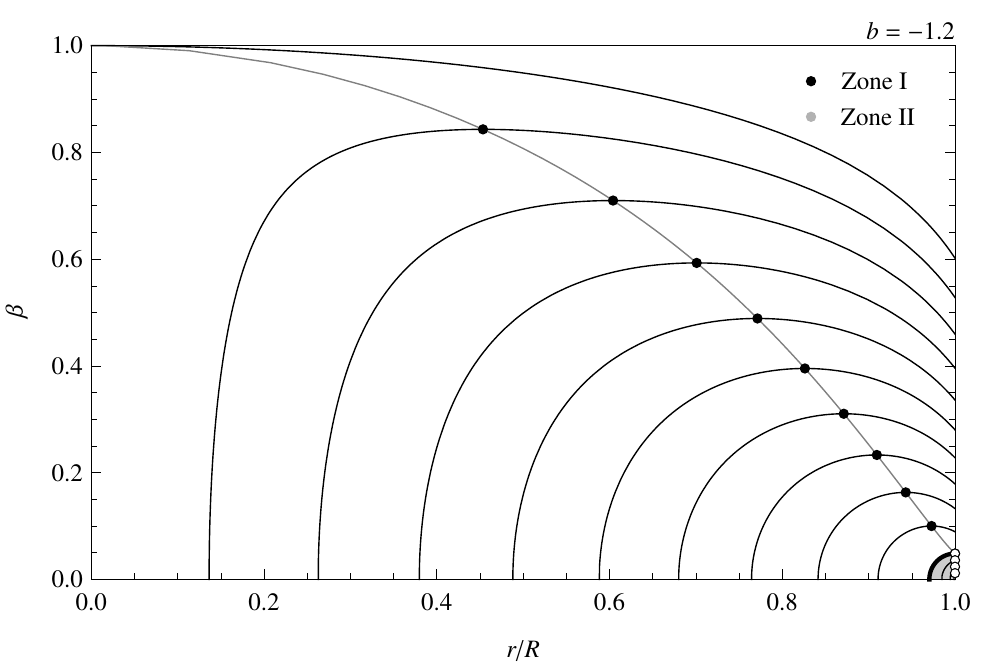}
         \includegraphics[width=.95\hsize,keepaspectratio=true]{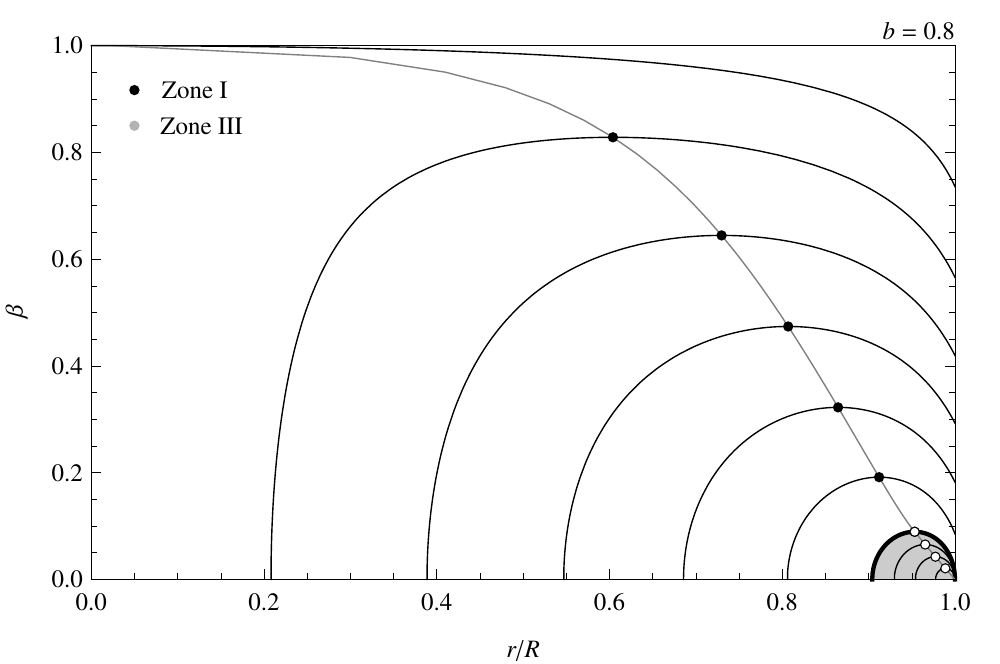}
      \end{center}
      \caption{Profiles of the local trapping factor $\beta (r/R, b)$ in the ECS with external spacetime of a black-hole type. They are constructed for fixed characteristic values of the brane parameter $b$ and sequences of the surface radius $R$ corresponding to the ECS. The upper figure represents the case of negative tidal charges ($b = -1.2$), while the lower figure represents the case of positive tidal charges ($b = 0.8$). The gray line connects the local maxima of the $\beta (r/R, b)$ profiles. The $\beta (r/R, b)$ profiles constructed for parameters belonging to different zones of the parameter space are separated by thick lines. In the regions of different zones the profiles are constructed for surface radii $R$ spaced by equal distances in the interval of maximal and minimal values of $R$ of ECS.}\label{FIGbetaA}
    \end{figure}
    \afterpage{\clearpage}
    \begin{figure}[ht]
      \begin{center}
         \includegraphics[width=.95\hsize,keepaspectratio=true]{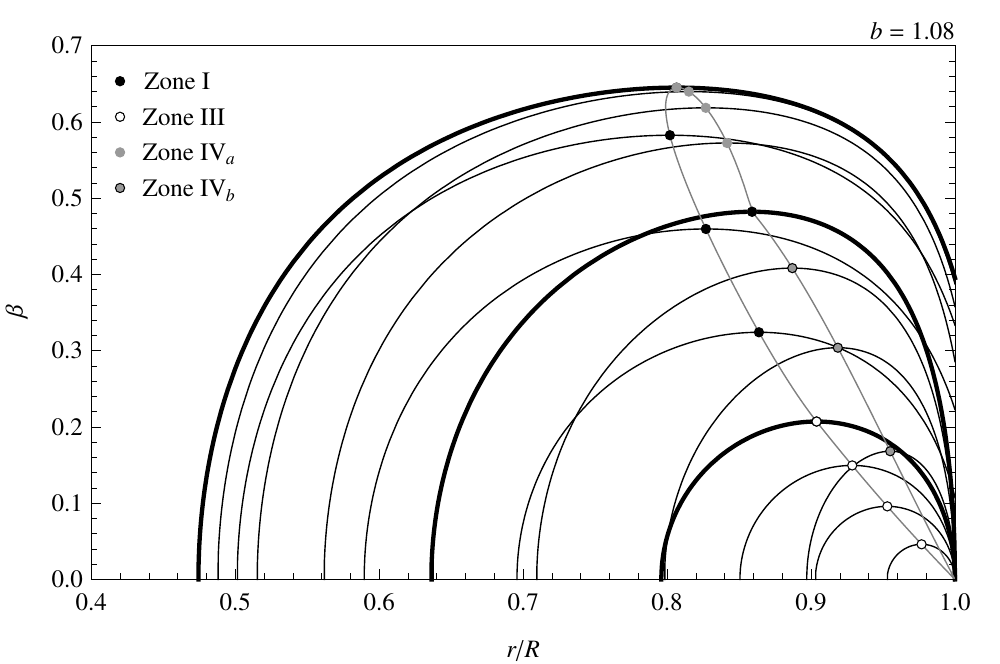}
         \includegraphics[width=.95\hsize,keepaspectratio=true]{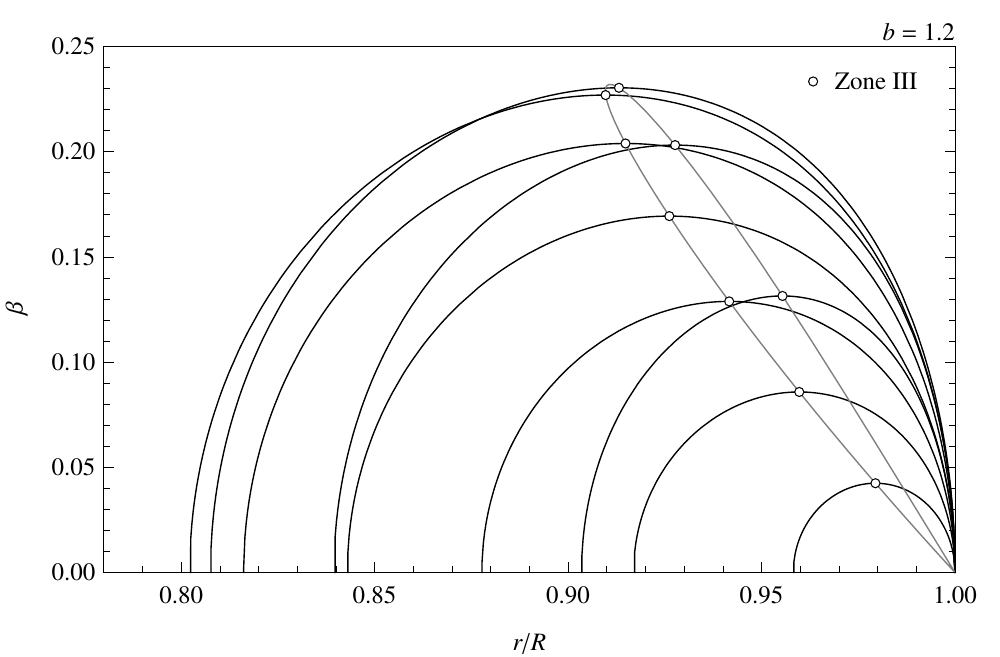}
      \end{center}
      \caption{Profiles of the local trapping factor $\beta (r/R, b)$ in the ECS with external spacetime of a naked-singularity type, constructed for fixed characteristic values of the brane parameter $b$ and sequences of the surface radius $R$ corresponding to the ECS. The upper figure corresponds to the case when two circular null geodesics of the external spacetime can exist ($b = 1.08$), while the lower figure corresponds only to the spacetimes where no circular null geodesic exists in the external spacetime ($b = 1.2$). The gray line connects the local maxima of the $\beta (r/R, b)$ profiles. The figures are constructed in the same way as those presented for the black-hole external spacetimes.}\label{FIGbetaB}
    \end{figure}

    Let $N_\mathrm{p}$, $N_\mathrm{e}$ and $N_\mathrm{b}$ denote, respectively, the number of produced, escaped and trapped neutrinos per unit time of an external static observer at infinity. In order to determine the global correction factors
    \begin{equation}
      \mathcal{E}(R, b)\equiv \frac{N_{\mathrm{e}}(R, b)}{N_{\mathrm{p}}(R, b)},\qquad
      \mathcal{B}(R, b)\equiv \frac{N_{\mathrm{b}}(R, b)}{N_{\mathrm{p}}(R, b)} = 1 - \mathcal{E}(R, b),
    \end{equation}
    it is necessary to introduce the local correction factor for trapped neutrinos at a given radius $r \in (r_{\mathrm{b(e)}}, R)$ (or $r \in (r_{\mathrm{b(i)}}, R)$, if relevant).

    In the general case of non-isotropic emission of neutrinos, the escaped-to-pro\-du\-ced ratio is given by the expression
    \begin{equation}
      \Omega_{\mathrm e}(\Psi_{\mathrm e}) = \int\limits_0^{\Psi_{\mathrm e}}\int\limits_0^{2\pi}p(\Psi)\sin\Psi \mathrm{d}\Psi \mathrm{d}\phi
    \end{equation}
    with $p(\Psi)$ being a directional function of the emission (scattering) process. Because of the assumption of isotropic emission of neutrinos in the frame of the static observers, the escaping solid angle $\Omega_\mathrm{e} (\Psi_\mathrm{e})$ determines fully the ratio of escaped-produced neutrinos and is given by
    \begin{equation}
      \Omega_{\mathrm e}(\Psi_{\mathrm e}) = \int\limits_0^{\Psi_{\mathrm e}}\int\limits_0^{2\pi}\sin\Psi \mathrm{d}\Psi\mathrm{d} \phi
      = 2 \pi (1-\cos\Psi_{\mathrm e}).
    \end{equation}
    The escaping correction factor
    \begin{equation}
      \epsilon(r, R, b) = \frac{\mathrm{d} N_{\mathrm{e}}(r, R, b)}{\mathrm{d} N_{\mathrm{p}}(r, R, b)}
        = \frac{2\Omega(\psi_{\mathrm{e}}(r, R, b))}{4\pi} = 1 - \cos\psi_{\mathrm{e}}(r, R, b),
    \end{equation}
    while the complementary factor for trapped neutrinos
    \begin{equation}
      \beta(r, R, b) = 1 - \epsilon(r, R, b)
        = \frac{\mathrm{d} N_{\mathrm{b}}(r, R, b)}{\mathrm{d} N_{\mathrm{p}}(r, R, b)} = \cos\psi_{\mathrm{e}}(r, R, B).
    \end{equation}
    Notice that we consider production and escaping rates at a given radius $r$, but the radius $R$ of the compact object enters the relation as it determines the escaping directional angle. The coefficient $\beta(r, R, b)$ determines local efficiency of the neutrino trapping, i.e., the ratio of the trapped and produced neutrinos at any given radius $r \in(r_{\mathrm{b(e)}}, R)$ or $r \in(r_{\mathrm{b(i)}}, R)$. Its profile is shown for several representative values of the tidal charge $b$ and related sequences of the surface radius $R$ in Figures~\ref{FIGbetaA}, \ref{FIGbetaB}.
    The local maxima of the function $\beta(r, R, b)$ (with $R$ and $b$ being fixed) are given by the condition $\partial \beta / \partial r = 0$ which is satisfied at radius $r = r_{\mathrm{c(i)}}$  implying coincidence with the radius of the stable circular null geodesic, as anticipated intuitively. For the Zone~II the maxima are naturally located at the surface radius. In  Figures~\ref{FIGbetaA}, \ref{FIGbetaB}, the maxima are depicted explicitly.

    We can see that the local trapping factor in any given ECS spacetime reaches its maximal values at the radius corresponding to the stable circular null geodesic and decreases as the radius falls to the centre of the ECS. Nevertheless, the trapping region can extend down to the region near the centre only for the most compact ECS with external spacetime of the black-hole type (for both negative and positive tidal charges), having surface radius close the minimal allowed surface radius. The maximum of the $\beta(r, R, b)$ profile then reaches value of $\beta \sim 1$ close to the centre, where the stable null circular geodesic is located in such spacetimes. On the other hand, for positively tidally charged ECS with external spacetime of the naked-singularity type, the $\beta(r, R, b)$ profile cannot reach the central part and its maximum is much smaller than $\beta = 1$.

  \subsection{Neutrino production rates}
    Generally, the neutrino production is a very complex process depending on detailed structure of an extremely compact object. We can express the locally defined neutrino production rate in the form
    \begin{equation}
      \mathcal{I}(r\{\mathcal{A}\}) = \frac{\mathrm{d} \mathcal{N}(r\{\mathcal{A}\})}{\mathrm{d} \tau(r)},
    \end{equation}
    where d$\mathcal{N}$ is the number of interactions at radius $r$, $\tau$ is the proper time of the static observer at the given $r$, $\{\mathcal{A}\}$ is the full set of quantities relevant for the production rate. We can write that
    \begin{equation}
      \mathrm{d} \mathcal{N}(r) = n(r)\Gamma(r)\mathrm{d} V(r),
    \end{equation}
    where $n(r)$, $\Gamma(r)$ and $\mathrm{d} V(r)$ are the number density of particles entering the neutrino production processes, the neutrino production rate and the proper volume element at the radius $r$, respectively. Both $n(r)$ and $\Gamma(r)$ are given by detailed structure of the extremely compact
    objects, $\mathrm{d} V(r)$ is given by the spacetime geometry.

    Here, considering the uniform energy density braneworld stars (for requirements of more realistic model see, e.g., \cite{Oest:2001:RAGtime2and3:,Web:1999:Pul:}), we shall assume the local production rate to be proportional to the energy density, i.e., we assume uniform production rate as measured by the local static observers; of course, from the point of view of static observers at infinity, the production rate will not be distributed uniformly. (According to \cite{Gle:1992:PHYSR4:,Gle:2000:CompactStars:,Hae-Zdu:1986:NATURE:}, such toy model could be reasonable good starting point for more realistic calculations.)

    In the internal spacetime we can thus write the local neutrino production rate in the form
    \begin{equation}
      \mathcal{I}(r) = \frac{\mathrm{d} \mathcal{N}}{\mathrm{d} \tau} \propto \rho = \mathrm{const}.
    \end{equation}
    The local neutrino production rate related to the distant static observers is then given by the relation including the time-delay factor
    \begin{equation}
      I = \frac{\mathrm{d} N}{\mathrm{d} t} = \mathcal{I} A(r, R, b).
    \end{equation}
    The number of neutrinos produced at a given radius in a proper volume $\mathrm{d} V$ per unit time of a distant static observer (that are governed by the internal metric coefficients) is given by the relation
    \begin{equation}
      \mathrm{d} N_{\mathrm{p}}(r, R, b) = I(r, R, b)\,\mathrm{d} V(r) = 4\pi\mathcal{I}\,A(r, R, b)\,B(r, R, b)\,r^{2}\,\mathrm{d} r.
    \end{equation}
    Integrating through whole the compact object (from $0$ to $R$), we arrive to the global neutrino production rate in the form
    \begin{equation}
      N_{\mathrm{p}}(R, b) = 4\pi\mathcal{I} \int_{0}^{R} A^{-}(r, R, b)B^{-}(r, R, b)r^2\,\mathrm{d} r.
    \end{equation}
    In an analogical way, we can give the expressions for the global rates of escaping and trapping of the produced neutrinos:
    \begin{eqnarray}
      N_{\mathrm{e}}(R,b) &=& 4\pi\mathcal{I}
      \int_{r_{\mathrm {b(e)}}}^{R} (1-\cos\psi_{\mathrm{e}}(r, R, b))A^{-}(r, R, b)B^{-}(r, R, b) r^{2}
      \,\mathrm{d} r + \nonumber \\
      & & + N_{\mathrm{p}}(r_{\mathrm {b(e)}}),\\
      N_{\mathrm{b}}(R,b) &=& 4\pi\mathcal{I} \int_{r_{\mathrm{b(e)}}}^{R} \cos\psi_{\mathrm{e}}(r, R, b)A^{-}(r, R, b)B^{-}(r, R, b) r^{2}\,\mathrm{d} r,
    \end{eqnarray}
    where $r_{\mathrm{b(e)}}$ is the radius given by equation~(\ref{EQrbe}) and $\cos\Psi_{\mathrm{e}}(r, R, b)$ is determined by equation~(\ref{EQsicose}). In ECS spacetimes with parameters belonging to the Zones where $r_{\mathrm{b(e)}}$ has not good meaning, its role takes $r_{\mathrm{b(i)}}$, while we replace $\cos\psi_{\mathrm{e}}$ for $\cos\psi_{\mathrm{i}}$. In such ECS spacetimes all bound neutrinos are the internal ones.

  \section{Efficiency of neutrino trapping}\label{SECeffnt}
    In order to characterize the trapping of neutrinos in extremely compact stars, we introduce some coefficients giving the efficiency of the trapping effect in connection to the total neutrino luminosity and the cooling process in the period of the evolution of the star corresponding to the geodetical motion of neutrinos.

  \subsection{Trapping coefficient of total neutrino luminosity}
    The influence of the trapping effect on the total neutrino luminosity of ECS can be appropriately given by the coefficient $\mathcal B_L$ relating the number of neutrinos produced inside the whole compact star during unit time of distant observers and the number of those produced neutrinos that will be captured by the extremely strong gravitational field of the star. The total luminosity trapping coefficient is therefore given by the relation
    \begin{equation}
        \mathcal{B}_\mathrm{L}(R,b) = \frac{\int_{r_{\mathrm{b(e)}}}^{R}A^{-}(r,R,b)B^{-}(r, R, b)\cos\psi_{\mathrm{e}}(r, R,b)
        \,r^{2}\,\mathrm{d} r}{\int_{0}^{R} A^{-}(r,R,b) B^{-}(r,R,b) r^{2}\,\mathrm{d} r},
      \end{equation}
    where
      \begin{equation}
        A^{-}B^{-}=\frac{3 R Z \left[(R-b) Y + R (2 b - 3 R) Z\right]}{Y \left[2 R (2 b - 3 R) Z - b Y\right]},
      \end{equation}
    and the complementary luminosity ``escaping'' coefficient is determined by the simple formula
      \begin{equation}
        \mathcal{E}_\mathrm{L}(R, b) = 1 - \mathcal{B}_\mathrm{L}(R, b).
      \end{equation}

    We can, moreover, define other global characteristic coefficients. For the ``internal'' neutrinos with motion restricted to the interior of the star, we introduce a coefficient
    \begin{equation}
      \mathcal{Q}_\mathrm{L}(R, b) = \frac{\int_{r_{\mathrm{b(i)}}}^{R}A^{-}(r, R, b)B^{-}(r, R, b)\cos\psi_{\mathrm{i}}(r, R, b)\,r^{2}\,\mathrm{d} r}{\int_{0}^{R} A^{-}(r,R,b) B^{-}(r,R,b) r^{2}\,\mathrm{d} r}
    \end{equation}
    and for the ``external'' neutrinos, we can use a complementary coefficient
    \begin{equation}
      \mathcal{X}_\mathrm{L} = \frac{N_{\mathrm{ext}}}{N_{\mathrm{p}}} = \mathcal{B}_\mathrm{L} - \mathcal{Q}_\mathrm{L}.
    \end{equation}

    \begin{figure}[t]
    \centering\includegraphics[width=.49\hsize,keepaspectratio=true]{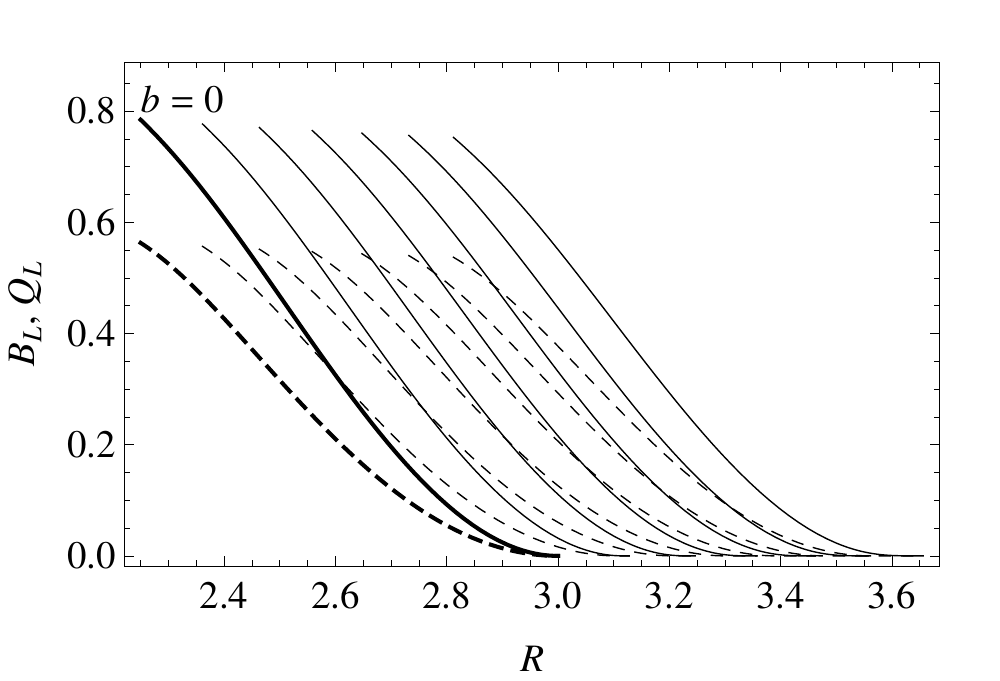}\includegraphics[width=.49\hsize,keepaspectratio=true]{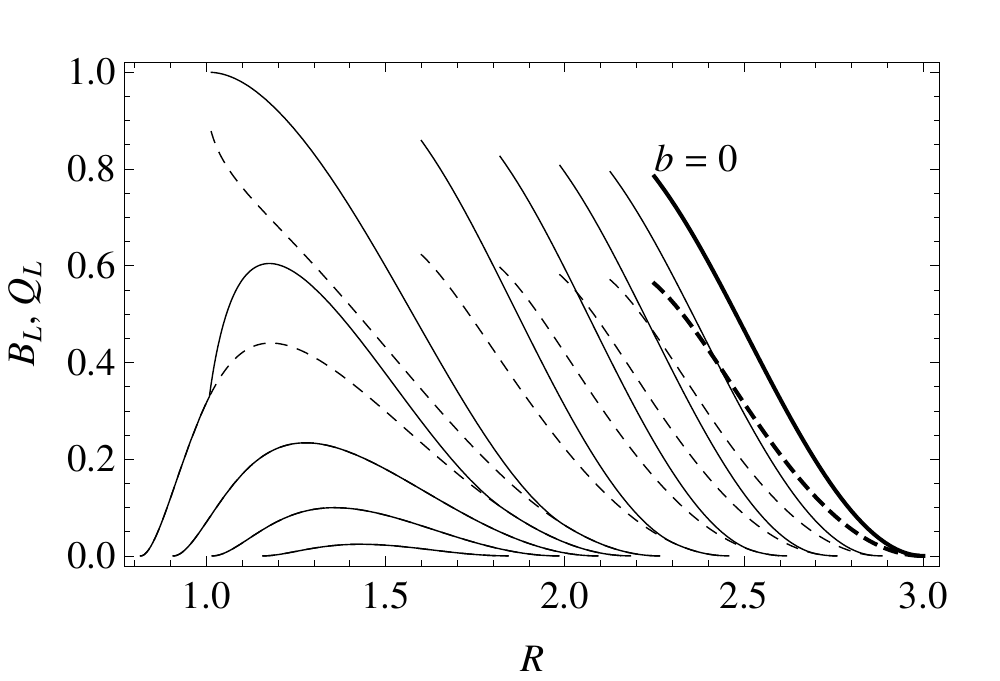}
    \caption{The total luminosity trapping coefficient $\mathcal{B}_\mathrm{L}(R, b)$ (full lines) and the internal luminosity trapping coefficient $\mathcal{Q}_\mathrm{L}(R, b)$ (dashed lines) as functions of the surface radius $R$ for fixed characteristic values of the tidal charge $b$. The left plot is given for negative tidal charges $-1.2 \leq  b\leq 0$, the right plot for positive tidal charges $0\leq b < 27/20$. The values of $b$ are equally spaced with interval of $0.2$ for $b < 1$ and $0.7$ for $b > 1$.}\label{FIGBQa}
    \end{figure}

    The results are illustrated for the coefficients $\mathcal{B}_\mathrm{L}(R, b)$, $\mathcal{Q}_\mathrm{L}(R, b)$ in  Figure~\ref{FIGBQa}. The behaviour of the total luminosity trapping factor is qualitatively the same for ECS with external spacetimes of the black-hole type for both negative and positive tidal charges, but it has qualitatively different character for the black-hole and naked-singularity ECS spacetimes. In the black-hole-type ECS it monotonously increases with decreasing surface radius for any fixed tidal charge (both negative and positive), while in the naked-singularity-type ECS spacetimes it reaches a maximum between the centre and the surface radius  $R$, and then it decreases to zero value for surface radius approaching the minimal value of $R$.

    For the black-hole-type ECS with negative tidal charges the maximum of the coefficient $\mathcal{B}_\mathrm{L}(R, b)$ slightly decreases with decreasing $b$, while for positively tidally charged ECS it increases with increasing $b$ approaching $\mathcal{B}_\mathrm{L}(R, b) = 1$ for $b \rightarrow 1$. We observe a similar behavior also for $\mathcal{Q}_\mathrm{L}(R, b)$, but the magnitude of this coefficient is smaller in comparison with $\mathcal{B}_\mathrm{L}(R, b)$. In the naked-singularity-type ECS (with positive tidal charges) the local maximum of $\mathcal{B}_\mathrm{L}(R, b)$ (and $\mathcal{Q}_\mathrm{L}(R, b)$) strongly decreases with $b$ increasing; of course, for ECS in the Zone~III of the space of parameters, the parameter $\mathcal{Q}_\mathrm{L}(R, b)$ is not defined.

  \subsection{Trapping coefficient of neutrino cooling process}
    The efficiency of the influence of neutrino trapping on the cooling process is most effectively described by the local coefficient of trapping $b_\mathrm{c}$ relating the trapped and produced neutrinos at a given radius of the star that is defined by the relation
    \begin{equation}
      b_\mathrm{c} (r, R, b) \equiv \beta (r, R, b).
    \end{equation}
    The local cooling coefficient is therefore given in Figures~\ref{FIGbetaA}, \ref{FIGbetaB} for appropriately chosen tidal charges $b$ and related sequences of the surface radius $R$. All the properties of the local trapping coefficient discussed above are thus relevant for the local cooling phenomena.

    \begin{figure}[t]
      \centering\includegraphics[width=.7\hsize,keepaspectratio=true]{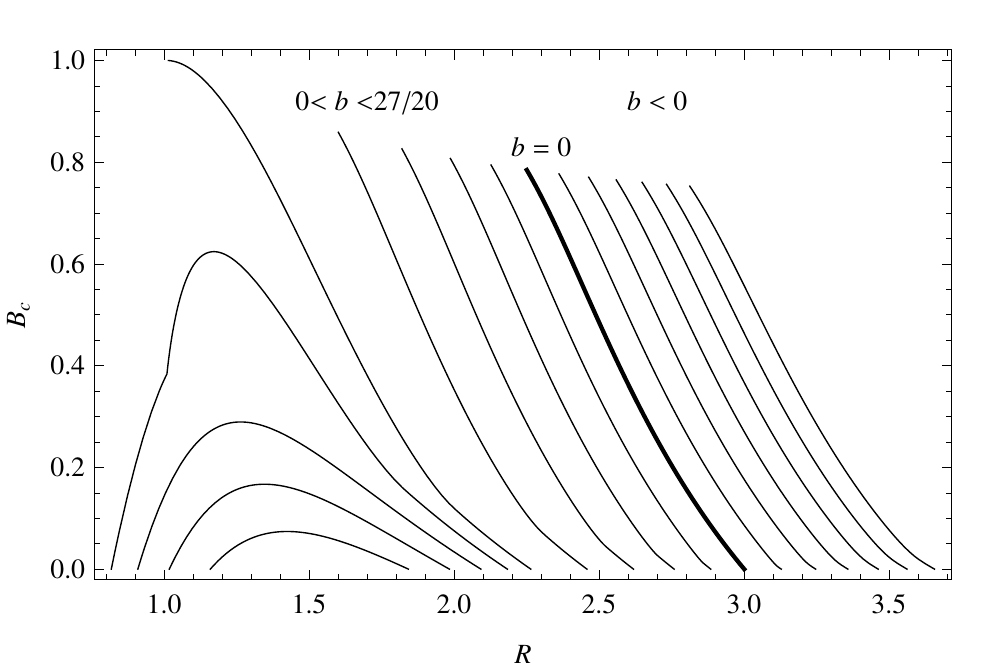}
      \caption{The global cooling coefficient related to trapping of neutrinos in the active zone of trapping $\mathcal{B}_\mathrm{C}(R, b)$. For tidal charge $b$ fixed for values with the same interval as in the previous figure and the full range of $R$ corresponding to the ECS is scanned.}\label{FIGBC}
    \end{figure}

    Further, the cooling process can be appropriately described in a complementary manner by a global coefficient for trapping, restricted to the ``active'' zone, where the trapping of neutrinos occurs. The global cooling coefficient is thus defined by the relation
    \begin{equation}
      \mathcal{B}_\mathrm{c}(R, b) \equiv \frac{\int_{r_\mathrm{b(e, i)}}^R A^{-}(r, R, b)B^{-}(r, R, b)\cos\psi_{\mathrm{e, i}}(r, R, b)\mathrm{d} r}
      {\int_{r_\mathrm{b(e, i)}}^R A^{-}(r, R, b)B^{-}(r, R, b) r^{2}\mathrm{d} r}.
    \end{equation}
    The global ``cooling'' coefficient $\mathcal{B}_\mathrm{c} (R, b)$
    is presented in Figure~\ref{FIGBC} --- for few characteristic values of the tidal charge $b$, both negative and positive, the full range of the ESC surface radius is scanned.

    The behaviour of the global cooling factor $\mathcal{B}_\mathrm{c}(R, b)$ is of the same character as for the total luminosity factor $\mathcal{B}_\mathrm{L}(R, b)$ in both black-hole and naked-singularity-type ECS and for both negative and positive tidal charges. Its magnitude is nearly twice higher than those of the total luminosity factor for the surface radii $R$ close to their maximal value, but it is nearly equal to the total luminosity factor as $R$ approaches its minimal value.

  \section{Conclusions}\label{SECconcl}
    We have demonstrated that the braneworld ECS can be separated into five classes corresponding to five Zones of the parameter space $(b-R)$ where trapping of neutrinos and their motion in the external field of the ECS are realized in qualitatively different way. There are fundamental differences of the trapping effect in the field of ECS with negative and positive tidal charges. The differences are related to both the surface radius of the ECS and the trapping coeffients of both kind --- for the total luminosity and the cooling process. Moreover, there is a strong difference of the trapping efficiency in the ECS with positive tidal charge when related to the black-hole spacetimes and the naked-singularity spacetimes. In the first case, the trapping is very efficient and the global trapping coefficient can approach the value of $1$ for the spacetimes with $b \rightarrow 1$, while in the second case the coefficients strongly decrease for tidal charges growing above the critical value of $b = 1$.

    There is an astrophysically very important effect related to the surface radius of the ECS with the negative and positive tidal charges. For $b < 0$, the surface radius of ECS can significantly overcome the limiting value of $R = 3$ valid for $b = 0$. On the other hand, for $b > 0$, the surface radius of ECS decreases under the standard value of $R = 3$ --- for ECS with $b \sim 1$, there is $R_\mathrm{max} \sim 2.3$ that is substantially lower than the values indicated by observations, and for $b > 1$ the ECS surface radius decreases further.

    In the ECS spacetimes with negative charge there is a crucial effect related to the increase of the surface radius of the ECS allowed for decreasing $b < 0$. For tidal charges slightly under the value of $b \sim -1$, the ECS surface radius approaches $R = 4$ strongly overcoming the limit of $R = 3$ corresponding to the standard Schwarzschild spacetimes and enters the region of values commonly accepted for observed neutron stars that can be lowered down to $R = 3.5$ \cite{Lat-Pra:2007:PhysRep:,Cot-Pae-Men:2002:NATURE:}. In such cases the coefficient of total luminosity can be of quite significant value for astrophysically realistic radii of ECS. For example, in the case of ECS with $R = 3.5$ and $b = -1.2$, there is $\mathcal{B}_\mathrm{L}(R, b) \sim 0.023$. The global cooling coefficient can approach even the value of $\mathcal{B}_\mathrm{c}(R, b) \sim 0.077$ under such conditions, indicating thus very strong and significant effects of trapping for braneworld negatively charged ECS, significantly overcoming the related values corresponding to the standard (tidally non-charged) spacetimes \cite{Stu-Tor-Hle-Urb:2009:}. Because the effect of trapping of neutrinos is cumulative, one can expect its relevance in realistic models of ECS to be strongly enhanced by negative tidal charges. Note that the conditions on the tidal charge magnitude implied by the high-frequency quasiperiodic oscillations observed in some low-mass-X-ray- binaries containing neutron or quark stars put the limit $b<1.2$ (assuming their mass $m \sim 2 M_{\odot}$) \cite{Kot-Stu-Tor:2008:CLASQG:}. This limit allows for sufficient increasing of the ECS surface radius with negative tidal charges enabling strong effects of neutrino trapping in the interior of such ECS with observationally acceptable radius. The expected lowering of the neutrino luminosity in observed neutron stars can be about $\sim 10\,\%$ and could be observationally tested.

    We can conclude that the negative tidal charges cause slight decrease of the efficiency of the trapping phenomena in the field of ECS, but they significantly shift these phenomena to the region of astrophysically relevant situations due to the strong increase of the surface radius of such ECS, shifting the radius to values acceptable observationally. On the other hand, the positive tidal charges generally enhance the trapping effects for the ECS spacetimes of the black-hole type, but the surface radius of such ECS is strongly shifted to the regions excluded observationally.

   \begin{acknowledgements}
     The present work was supported by the Czech grants MSM~4781305903, LC~06014, GA\v{C}R~205/09/H033 and the internal grant SGS/2/2010. One of the authors (Z.\,S.) would like to express his gratitude to the Czech Committee for collaboration with CERN and the Theory Division of CERN for perfect hospitality.
   \end{acknowledgements}


\begin{thebibliography}{99}
     \bibitem{Abdu-Ahme:2010:PHYSR4}
       Abdujabbarov, A., Ahmedov, B.:
       Test particle motion around a black hole in a braneworld.
       Phys. Rev. D \textbf{81}, 044022 (2010)

     \bibitem{Abr-etal:1997:CLAQG:}
       Abramowicz, M.A., Anderson, N., Bruni, M., Ghosh, P., Sonego, S.:
       Gravitational waves from ultracompact stars: the optical geometry view of trapped modes.
       Class. Quantum Grav. \textbf{14}, L189--L194 (1997)

     \bibitem{Abr-Mil-Stu:1993:PHYSR4:}
       Abramowicz, M.A., Miller, J.C., Stuchl{\'{\i}}k, Z.:
       Concept of radius of gyration in general relativity.
       Phys. Rev. D \textbf{47} 1440--1447 (1993)

     \bibitem{Abr-Pra:1990:}
       Abramowicz, M.A., Prasanna, A.R.:
       Centrifugal Force Reversal Near a Schwarzschild Black-Hole.
       Mon. Not. R. Astron. Soc. \textbf{245}, 720 (1990)

     \bibitem{Ali-Gum:2005:}
       Aliev, A.N., G{\"u}mr{\"u}k{\c c}{\"u}o{\u g}lu, A.E.:
       Charged rotating black holes on a 3-brane.
       Phys. Rev. D \textbf{71}, 104027 (2005)

     \bibitem{Ali-Tal:2009:}
       Aliev, A.N., Talazan, P.:
       Gravitational effects of rotating braneworld black holes.
       Phys. Rev. D \textbf{80}, 044023 (2009)

     \bibitem{Ark-Dim-Dva:1998:}
       Arkani-Hamed, N., Dimopoulos, S., Dvali, G.:
       The hierarchy problem and new dimensions at a millimeter.
       Phys. Lett. B \textbf{429}, 263--272 (1998)

     \bibitem{Bah-Lyn-Sel:1990:ApJ:}
       Bahcall, S., Lynn, B.W., Selipsky, S.~B.:
       New models for neutron stars.
       Astrophys. J. \textbf{362}, 251--255 (1990)

     \bibitem{Boh-Ris-Har-Lob:2010:CLASQG}
       B{\"o}hmer, C.G., De Risi, G., Harko, T., Lobo, F.S.N.:
       Classical tests of general relativity in brane world models.
       Class. Quantum Grav. \textbf{27}, 185013 (2010)

     \bibitem{Boh-Har-Lob:2008:CLAQG:}
       B{\"o}hmer, C.G., Harko, T., Lobo, F.S.N.:
       Solar system tests of brane world models.
       Class. Quantum Grav. \textbf{25}, 045015 (2008)

     \bibitem{Bin-Nun:2010:PHYSR4:b}
       Bin-Nun, A.Y.:
       Relativistic images in Randall-Sundrum II braneworld lensing.
       Phys. Rev. D \textbf{81}, 123011 (2010)

     \bibitem{Bin-Nun:2010:PHYSR4:a}
       Bin-Nun, A.Y.:
       Gravitational lensing of stars orbiting Sgr A* as a probe of the black hole metric in the Galactic center.
       Phys. Rev. D \textbf{82}, 064009 (2010)

     \bibitem{Cot-Pae-Men:2002:NATURE:}
       Cottam, J., Paerls, F., Mendez, M.:
       Gravitationally redshifted absorption lines in the X-ray burst spectra of a neutron star.
       Nature \textbf{420}, 51--54 (2002)

     \bibitem{Dad-Maar-Pap-Rez:2000:PHYSR4:}
       Dadhich, N., Maartens, R., Papadopoulos, P., Rezania, V.:
       Black holes on the brane.
       Phys. Lett. B \textbf{487}, 1--6 (2000)

     \bibitem{Dim-Lan:2001:}
       Dimopoulos, S., Landsberg, G.:
       Black Holes at the Large Hadron Collider.
       Phys. Rev. Lett. \textbf{87}, 161602 (2001)

     \bibitem{Ger-Maar:2001:}
       Germani, C., Maartens, R.:
       Stars in the braneworld.
       Phys. Rev. D \textbf{64}, 124010 (2001)

     \bibitem{Gle:1992:PHYSR4:}
       Glendenning, N.K.:
       First-order phase transitions with more than one conserved charge: Consequences for neutron stars.
       Phys. Rev. D \textbf{46}, 1274--1287 (1992)

     \bibitem{Gle:2000:CompactStars:}
       Glendenning, N.K.:
       Compact Stars: Nuclear Physics, Particle Physics, and General Relativity.
       Springer, New York (2000)

     \bibitem{Hae-Zdu:1986:NATURE:}
       Haensel, P., Zdunik, J.L.:
       A submillisecond pulsar and the equation of state of dense matter.
       Nature \textbf{340}, 617--619 (1986)

     \bibitem{Hle-Stu-Mra:2004:RAGtime4and5:CrossRef}
       Hled\'{\i}k, S., Stuchl\'{\i}k, Z., Mr\'{a}zov\'{a}, K.:
       Comparison of general relativistic polytropic and adiabatic fluid spheres with a~repulsive cosmological constant.
       In: Hled\'{\i}k, S., Stuchl\'{\i}k, Z. (eds.) Proceedings of RAGtime 4/5: Workshops on black holes and neutron stars, Opava, 14--16/13--15 October 2002/03, pp.~75--89. Silesian University in Opava, Opava (2004)

     \bibitem{Kot-Stu-Tor:2008:CLASQG:}
       Kotrlov{\'a}, A., Stuchl{\'{\i}}k, Z., T{\"o}r{\"o}k, G.:
       Quasiperiodic oscillations in a strong gravitational field around neutron stars testing braneworld models.
       Class. Quantum Grav. \textbf{25}, 225016 (2008)

     \bibitem{Lat-Pra:2004:SCIENCE:}
       Lattimer, J., Prakash, M.:
       The physics of neutron stars.
       Science \textbf{304}, 536--542 (2004)

     \bibitem{Lat-Pra:2007:PhysRep:}
       Lattimer, J., Prakash, M.:
       Neutron star observations: Prognosis on Equation of state constraints.
       Phys. Rep. \textbf{442}, 109--165 (2007)

     \bibitem{Maar:2004:}
       Maartens, R.:
       Brane-World Gravity.
       Living Rev. Relativity \textbf{7}, 7 (2004)

     \bibitem{Mallick-etal:2009:arXiv:0905.3605:}
       Mallick, R., Bhattacharyya, A., Ghosh, S.K., Raha, S.:
       General Relativistic effect on the energy deposition rate for neutrino pair annihilation above the equatorial plane along the symmetry axis near a rotating neutron star.
       ArXiv e-prints 0905.3605v2 [astro-ph.HE] (2009)

     \bibitem{Mam-Hak-Toj:2010:MPLA:}
       Mamadjanov, A.I., Hakimov, A.A., Tojiev, S.R.:
       Quantum Interference Effects in Spacetime of Slowly Rotating Compact Objects in Braneworld.
       Mod. Phys. Lett. A \textbf{25}, 243--256 (2010)

     \bibitem{Mil-Sha-Nol:1998:MNRAS:}
       Miller, J., Shahbaz, T., Nolan, L. A.:
       Are Q-stars a serious threat for stellar-mass black hole candidates?
       Mon. Not. R. Astron. Soc. \textbf{294}, L25--L29 (1998)

     \bibitem{Mis-Tho-Whe:1973:Gra:}
       Misner, C.W., Thorne, K.S., Wheeler, J.A.:
       Gravitation.
       San Francisco, W. H. Freeman (1973)

     \bibitem{Mor-Ahme-Abdu-Mam:2010:ASS:}
       Morozova, V.S., Ahmedov, B.J., Abdujabbarov, A.A., Mamadjanov, A.I.:
       Plasma magnetosphere of rotating magnetized neutron star in the braneworld.
       Astrophys. \&{} Space Sci. \textbf{330}, 257--266 (2010)

     \bibitem{Nil-Cla:2000:GRRelStarsPolyEOS:}
       Nilsson, U.S., Ugla, C.:
       General Relativistic Stars: Polytropic Equations of State.
       Annals of Physics \textbf{286}, 292--319 (2000)

     \bibitem{Oest:2001:RAGtime2and3:}
       {\O}stgaard, E.:
       Internal structure of neutron stars.
       In: Hled{\'{\i}}k, S., Stuchl{\'{\i}}k, Z. (eds.) Proceedings of RAGtime 2/3: Workshops on black holes and neutron stars, Opava, 11--13/8--10 October 2000/01,  pp.~73--102. Silesian University in Opava, Opava (2001)

     \bibitem{Ran-Sun:1999:}
       Randall, L., Sundrum, R.:
       An Alternative to Compactification.
       Phys. Rev. Lett. \textbf{83}, 4690--4693 (1999)

     \bibitem{Sche-Stu:2009:b}
       Schee, J., Stuchl{\'{\i}}k, Z.:
       Profiles of emission lines generated by rings orbiting braneworld Kerr black holes.
       Gen. Relativ. Gravit. \textbf{41}, 1795--1818 (2009)

     \bibitem{Sche-Stu:2009:a}
       Schee, J., Stuchl{\'{\i}}k, Z.:
       Optical Phenomena in the Field of Braneworld Kerr Black Holes.
       Int. J. Mod. Phys. D \textbf{18}, 983--1024 (2009)

     \bibitem{Schw:1916:SITBA:}
       Schwarzschild, K.:
       \"{U}ber das Gravitationsfeld einer Kugel aus inkompressibler Fl\"{u}ssigkeit nach der Einsteinschen Theorie.
       Sitzungsber. K. Preuss. Akad. Wiss., Phys.--Math. Kl. 424--434 (1916)

     \bibitem{Sha-Teu:1983:BHWDNS:}
       Shapiro, S.L., Teukolsky, S.A.:
       Black Holes, White Dwarfs and Neutron Stars: The Physics of Compact Objects.
       Wiley--VCH, New York (1983)

     \bibitem{Shi-Mae-Sas:1999:}
       Shiromizu, T., Maeda, K.-I., Sasaki, M.:
       The Einstein equations on the 3-brane world.
       Phys. Rev. D \textbf{62}, 024012 (2000)

     \bibitem{Stu:1990:}
       Stuchl{\'{\i}}k, Z.:
       Note on the properties of the Schwarzschild-de-Sitter spacetime.
       Bull. Astronom. Inst. Czechoslovakia \textbf{41}, 341--343 (1990)

     \bibitem{Stu:2000:ACTPS2:}
       Stuchl\'{\i}k, Z.:
       Spherically symmetric static configurations of uniform density in spacetimes with a non-zero cosmological constant.
       Acta Phys. Slovaca, \textbf{50}, 219--228 (2000)

     \bibitem{Stu-Hle-Jur:2000:}
       Stuchl{\'{\i}}k, Z., Hled{\'{\i}}k, S., Jur{\'a}\v{n}, J.:
       Optical reference geometry of Kerr-Newman spacetimes.
       Class. Quantum Grav. \textbf{17}, 2691--2718 (2000)

     \bibitem{Stu-etal:2001:PHYSR4:}
       Stuchl\'{\i}k, Z., Hled\'{\i}k, S., \v{S}olt\'{e}s, J., {\O}stgaard, E.:
       Null geodesics and embedding diagrams of the interior Schwarzschild--de~Sitter spacetimes with uniform density.
       Phys. Rev. D \textbf{64}, 044004 (2001)

     \bibitem{Stu-Kot:2009:}
       Stuchl{\'{\i}}k, Z., Kotrlov{\'a}, A.:
       Orbital resonances in discs around braneworld Kerr black holes.
       Gen. Relativ. Gravit. \textbf{41}, 1305--1343 (2009)

     \bibitem{Stu-Schee:2010:CLAQG:}
       Stuchl{\'{\i}}k, Z., Schee, J.:
       Appearance of Keplerian discs orbiting Kerr superspinars.
       Class. Quantum Grav. \textbf{27}, 215017 (2010)

     \bibitem{Stu-Tor-Hle-Urb:2009:}
       Stuchl{\'{\i}}k, Z., T{\"o}r{\"o}k, G., Hled{\'{\i}}k, S., Urbanec, M.:
       Neutrino trapping in extremely compact objects: I. Efficiency of trapping in the internal Schwarzschild spacetimes.
       Class. Quantum Grav. \textbf{26}, 035003 (2009)

     \bibitem{Web:1999:Pul:}
       Weber, F.:
       Pulsars as Astrophysical Laboratories for Nuclear and Particle Physics.
       Taylor \&{} Francis, London (1999)

     \bibitem{Web-Gle:1992:ASTRJ2:}
       Weber, F., Glendenning, N.K.:
       Application of the improved Hartle method for the construction of general relativistic rotating neutron star models.
       Astrophys. J. \textbf{390}, 541 (1992)
   \end{thebibliography}
\end{document}